\documentclass[11pt]{article}

\usepackage[final]{acl}

\usepackage{times}
\usepackage{latexsym}

\usepackage[T1]{fontenc}

\usepackage[utf8]{inputenc}
\usepackage{booktabs}
\usepackage{multirow}
\usepackage{amsmath}
\usepackage{microtype}

\usepackage{inconsolata}

\usepackage{graphicx}

\usepackage{xspace}
\usepackage[table]{xcolor}
\usepackage{enumitem}

\title{\ours: Supporting Open-ended Teamwork with Multi-Agent Systems}

\newcommand{\aspace}{\hspace{1em}}

\author{%
    \textbf{Jiale Liu}\textsuperscript{1} \aspace
    \textbf{Victor S. Bursztyn}\textsuperscript{2} \aspace
    \textbf{Lin Ai}\textsuperscript{3} \aspace
    \textbf{Haoliang Wang}\textsuperscript{2} \aspace\\
    \textbf{Sunav Choudhary}\textsuperscript{2} \aspace
    \textbf{Saayan Mitra}\textsuperscript{2} \aspace
    \textbf{Qingyun Wu}\textsuperscript{1,4} \aspace\\
    \textsuperscript{1}{}Pennsylvania State University\aspace \textsuperscript{2}{}Adobe Research\aspace \textsuperscript{3}Columbia University \aspace \textsuperscript{4}{}AG2ai, Inc.\\
    \texttt{\{jiale.liu, qingyun.wu\}@psu.edu} \\
    \texttt{lin.ai@cs.columbia.edu} \\
    \texttt{\{victor.bursztyn, haowang, schoudha, smitra\}@adobe.com}
}

\newcommand{\ours}{TeamFusion\xspace}

\newcommand{\result}[2]{%
  \ensuremath{#1_{\scriptstyle #2}}%
}
\newcommand{\bestresult}[2]{
  \ensuremath{\mathbf{#1}_{\scriptstyle #2}}%
}

\begin{document}
\maketitle
\begin{abstract}
In open-ended domains, teams must reconcile diverse viewpoints to produce strong deliverables. Answer aggregation approaches commonly used in closed domains are ill-suited to this setting, as they tend to suppress minority perspectives rather than resolve underlying disagreements.
We present \ours, a multi-agent system designed to support teamwork in open-ended domains by: 1. Instantiating a proxy agent for each team member conditioned on their expressed preferences; 2. Conducting a structured discussion to surface agreements and disagreements; and 3. Synthesizing more consensus-oriented deliverables that feed into new iterations of discussion and refinement.
We evaluate \ours on two teamwork tasks where team members can assess how well their individual views are represented in team decisions and how consensually strong the final deliverables are, finding that it outperforms direct aggregation baselines across metrics, tasks, and team configurations.


\end{abstract}

\section{Introduction}
Many group decisions are open-ended: there is no single correct answer, but multiple plausible options that trade off values, constraints, and risk~\citep{black1948rationale,kiesler1992group,kraemer1988computer}. In these settings, success is not “matching the gold label,” but producing a deliverable that group participants recognize as reflecting their distinct preferences and rationales~\citep{fisher1970decision}. However, arriving at such a deliverable is expensive: teams must surface hidden assumptions, productively discuss disagreements, and negotiate acceptable trade-offs, which create communication bottlenecks and high costs at scale~\citep{ROMNEY202533,Rogelberg2006NotAM}.


\begin{figure}[t]
  \centering
  \includegraphics[width=1\columnwidth]{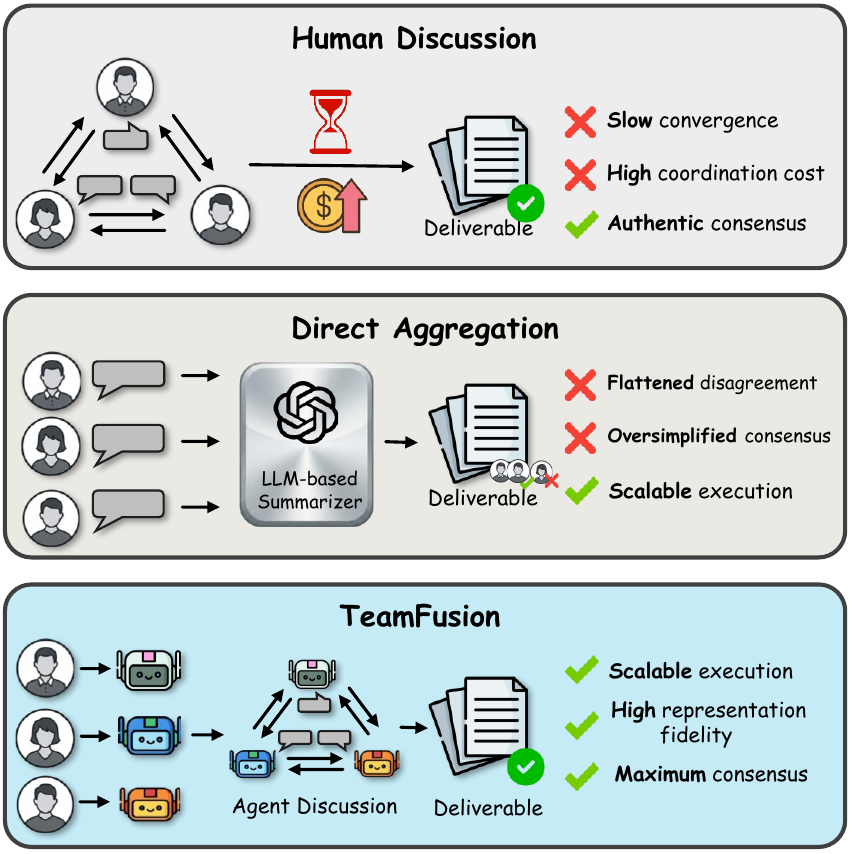}
  \caption{Illustration of \ours versus baselines. While human discussion is slow and direct aggregation loses nuance, \ours leverages agent-based discussion to combine fast execution with the high representation fidelity and maximum consensus.}
\end{figure}

Large language models (LLMs) appear promising for decision support because they can digest large amounts of text and draft deliverables that people can critique and revise~\citep{zhang2025systematic,naveed2025comprehensive}. Yet many existing LLM usages in group settings still follow direct aggregation: concatenate inputs and generate a single recommendation~\citep{bhaskar-etal-2023-prompted,li-etal-2024-sentiment,li-etal-2023-summarizing}, or collapse rationales into an ``average'' feedback~\citep{zhu2025can,huang-etal-2023-examining}. This approach is ill-suited for open-ended teamwork for two reasons. First, a single-shot aggregate can be hard to audit and may introduce ungrounded claims, which is problematic when the deliverable must be attributable to participants’ stated reasons~\citep{huang2025survey,parcalabescu2024measuring,liu2023trustworthy,yu2025survey}. Second, direct aggregation can suppress disagreements instead of attempting to resolve them in a refined deliverable~\citep{zhu2025can,laban2023summedits,zhang2024comprehensive,wang2023survey}. In other words, teamwork is not only about capturing shared commonsense, but also discussing local disagreements to make deliverables well-rounded overall. 

This gap motivates our work: \emph{how can we generate deliverables that preserve diverse individual viewpoints while helping teams to converge?} We argue that closing this gap requires modeling the iterative consensus-seeking process. In open-ended decisions, key information emerges when perspectives respond to one another: participants clarify opinions, challenge missing cases, and refine proposals in light of others’ objections. A system that skips this step must implicitly guess the structure of disagreement from raw text, which is precisely where viewpoint erasure occurs.

We introduce \ours, a general multi-agent framework for open-ended teamwork support. \ours (i) instantiates a proxy agent for each team member, conditioned on their expressed preferences; (ii) runs a structured discussion to make agreements and disagreements explicit; and (iii) synthesizes the discussion into an editable deliverable that records trade-offs and supporting reasons. Our central hypothesis is that explicitly modeling team members and their interaction yields deliverables that are both more representative of diverse viewpoints and more useful for decision-making than direct aggregation.

We evaluate \ours on two teamwork tasks where team members can judge how well their individual views are represented in team decisions and how consensually strong the final deliverables are. The results indicate that \ours outperforms baselines across metrics, tasks, backbone models, and team configurations.
Our contributions are:
\vspace{-4pt}
\begin{enumerate}[leftmargin=*,noitemsep]
    \item We propose \ours, a framework for open-ended decision support in teams. By modeling the process of consensus-seeking, \ours synthesizes deliverables that cover wider viewpoints while driving convergence.
    \item We propose a scalable, human-in-the-loop evaluation protocol for open-ended team tasks. By decoupling preference collection from interaction, our protocol overcomes the logistical bottlenecks of synchronous team studies, allowing for rigorous, large-scale evaluation of AI tools with professional domain experts.
    \item Our results generalize across text and multimodal tasks, as well as teams of different sizes, showing that structured agent interaction yields higher-quality deliverables.
\end{enumerate}


\section{Related Work}

\subsection{Multi-agent Systems}
Recent work has explored ``societies'' of LLM with agents that interact via structured dialogue~\citep{piatti2024cooperate,park2023generative}.
General-purpose orchestration frameworks such as AutoGen~\citep{wu2024autogen}, MetaGPT~\citep{hong2023metagpt} and LangChain~\citep{chase_langchain_2022} make it easier to construct multi-agent systems via role assignment, tool use, and customizable interaction protocols.
Within this broader trend, multi-agent debate has emerged as a simple but effective recipe: multiple model instances propose answers, critique one another, refine a final response~\citep{du2023improving,chan2023chateval,liang2024encouraging}. Extensive work has shown that by orchestrating and integrating agent responses, the system can generate outputs that are more factual~\citep{du2023improving,chern2024combating,kim2024mdagents}, creative~\citep{liang2024encouraging,hu2025debate}, and functionally correct~\citep{suncorex,zhang2025debate4math,song2024adaptive,zhang2024offline}.
Whereas debate frameworks primarily optimize for correctness or factuality, our focus is to use structured discussion to find and expand agreements, disagreements, and trade-offs in open-ended decisions.

\subsection{LLMs for Group Consensus}
Developing systems for group consensus has been a long-reaching question in NLP, with pre-LLM work building meeting corpora and identifying decision-related dialogue to support teams’ shared understanding~\citep{10.1007/11677482_3,shriberg-etal-2004-icsi,orwig1997multi}. With the advent and prevalence of LLMs, recent work increasingly leverages the model as a facilitator that steers deliberation: \cite{doi:10.1126/science.adq2852} proposed ``Habermas Machine,'' an LLM mediator that can help small groups find common ground in democratic deliberation, while structured conversational interventions can counter group think and improve how teams scrutinize AI advice during collective decisions~\citep{chiang2024enhancing}, and prompt-tuned mediation strategies can de-escalate or reframe online conflict toward agreement~\citep{10.1145/3613904.3642322}.
Building on this emerging view of LLMs as facilitators, \ours supports convergence of open-ended teamwork by representing each participant with a conditioned proxy agent, orchestrating a structured multi-party discussion, and synthesizing the discussion into a deliverable, iteratively.




\begin{figure*}[t]
  \centering
  \includegraphics[width=0.87\textwidth]{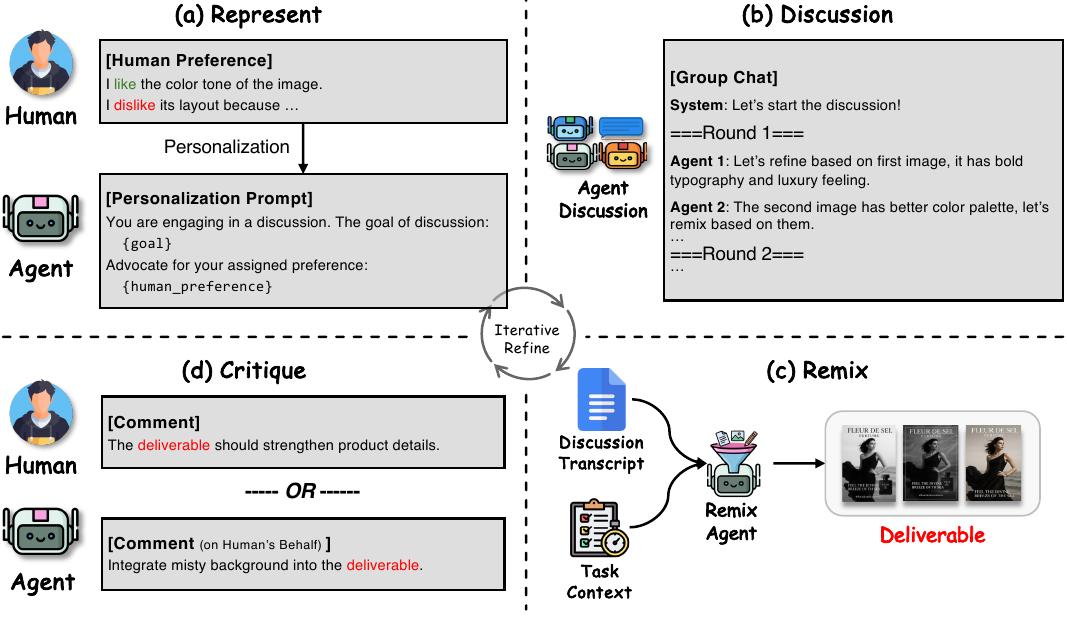}
  \caption{The overview of the \ours framework. It consists of four phases: (1) Represent: We extract human preference labels as agents; (2) Discussion: The agents abstracted from human preference engage in a structured discussion; (3) Remix: The discussion transcript along with task context are remixed into a final deliverable used directly for downstream decision.; (4) Critique and Refine: The agent or human leave critiques based on generated deliverable, and the system iterates again on improving the deliverable.}
  \label{fig:teamfusion_main}
\end{figure*}

\section{\ours Framework}
\label{sec:methods}


\paragraph{Problem setup.}
We study \emph{open-ended} teamwork, where the goal is to produce a deliverable that (i) preserves distinct viewpoints and constraints and (ii) helps a team move toward an acceptable outcome.
Given a task context $c$ and a set of $N$ team members $\{u_1,\ldots,u_N\}$, each providing task-specific preference $E_i$, \ours outputs a deliverable $y$ intended to be directly usable.

\paragraph{Overview}
As shown in Figure~\ref{fig:teamfusion_main}, \ours consists of four phases: (1) we instantiate one proxy agent per participant from their preferences; (2) proxy agents engage in a structured group discussion; (3) a remix phase converts the discussion into an editable deliverable, and (4) the system iterates this loop to refine the deliverable.

\subsection{Represent}

For each participant $u_i$, we create a proxy agent $a_i$ designed to argue from $u_i$'s perspective during the group discussion.
Training a separate model per participant is impractical in realistic settings: per-user data are sparse, training is computationally expensive, and models would quickly become obsolete as preferences shift. Instead, we adopt an in-context personalization approach.
Concretely, we encode the participant's evidence $E_i$ into a structured system prompt $\pi_i$ that specifies: (i) the agent's role and collaborative objective, (ii) domain and communication constraints, and (iii) the participant-specific preferences.
Our goal is not to fully model a participant's identity, but to ensure the agent's contributions are recognizably aligned with that participant's expressed perspectives.


\subsection{Discuss}
\label{debate}


\paragraph{Agent Roles}
One \ours run consists of $N$ participant proxy agents:
$A=\{a_1,\ldots,a_N\}$.
Proxy agents contribute proposals and critiques from their participant's perspective.

\paragraph{Conversation State}
The discussion proceeds in a shared group-chat environment. At step $t$, the controller maintains a message history $H_t=[m_1,\ldots,m_t]$, where each message $m_k=(\textit{name},\textit{content})$.
Agents do not hold additional private state. At each turn, the controller selects a speaker $a \in A$ and prompts the underlying LLM with the agent's system prompt $\pi_a$ and the current history $H_t$.
This ensures that all agents reason over the same dialogue context.

\paragraph{Turn-Taking Protocol}
We adopt a simple but effective round-robin protocol inspired by the classic divergence–convergence model of creative processes~\citep{acar2019divergent,runco2012divergent} and nominal group technique~\citep{10.1016/S0167-9236(00)00073-7}, giving each proxy agent a fixed number of speaking turns and cycling deterministically through agents to ensure equal opportunities to contribute. On a proxy turn, $a_i$ receives $H_t$ and is instructed to respond in light of its participant evidence and advance the discussion toward a recommendation. The discussion ends once all proxy agents exhaust their allotted turns.


\subsection{Remix}
\label{remix}

After the debate concludes, \ours converts the accumulated discussion into a final deliverable for the open-ended task at hand. A remixing agent takes as input the original task context $c$ and the full discussion history $H_T$, and produces a deliverable that combines all reasoning over disagreements and points of convergence. The remixed deliverable is intended to be directly consumable by humans experts on the task. Implementation details for each task are provided in Appendix~\ref{sec:remix_detail}.

\subsection{Iterative Refinement}
\label{iter}

\ours can be applied once to obtain a single deliverable, or used iteratively to gradually refine outputs.
In the iterative setting, the deliverable from one round is treated as a new proposal that re-enters the discussion: proxy agents are given access to the updated deliverable alongside the original context and asked to critique and build upon it in a subsequent discussion.
This refinement loop allows the system to successively narrow in on options that better reflect surfaced preferences and rationales.

\section{Task 1: Civic Comment Synthesis}
We begin by evaluating \ours in a civic decision-support setting, where a small group must turn diverse free-form public comments into a deliverable that can inform downstream action.

\subsection{Task Introduction}

Given a policy-relevant question and a set of participant comments, the system produces a concise summary intended to serve as a deliverable: it should capture the range of perspectives and the reasons behind them, rather than collapsing the group into a single averaged voice.

\subsection{Experiment Protocol}
\paragraph{Experiment Data}

We experiment on DeliberationBank~\citep{zhu2025can}, a benchmark containing U.S.-based public opinion comments spanning ten questions about technology, social media, and public policy.

\begin{table*}[t]
  \centering
  \small
  \setlength{\tabcolsep}{4pt}
  \begin{tabular}{l|l*{4}{c}*{4}{c}}
    \toprule
    \addlinespace[1pt]
    \midrule
    \multirow{2}{*}{Model} & \multirow{2}{*}{Method} &
      \multicolumn{4}{c}{OpenQA} & \multicolumn{4}{c}{BinaryQA} \\
    \cmidrule(lr){3-6} \cmidrule(lr){7-10}
    & &
    Represent. & Inform. & Neutral. & Policy &
    Represent. & Inform. & Neutral. & Policy \\
    \midrule
    \multirow{5}{*}{Llama-3.3-70B} &
      Direct &
        \result{.586}{.006} & \result{.537}{.006} & \result{.580}{.006} & \result{.574}{.007} &
        \result{.595}{.007} & \result{.541}{.006} & \result{.590}{.007} & \result{.542}{.007} \\
    & CoT &
        \result{.568}{.007} & \result{.524}{.006} & \result{.570}{.006} & \result{.561}{.007} &
        \result{.578}{.007} & \result{.530}{.006} & \result{.580}{.006} & \result{.528}{.007} \\
    & Self-Refine &
    \result{.577}{.007} & \result{.536}{.006} & \result{.566}{.006} & \result{.570}{.007} &
    \result{.599}{.007} & \result{.545}{.006} & \result{.587}{.007} & \result{.547}{.007} \\
    & MAD &
    \result{.588}{.007} & \result{.544}{.006} & \result{.572}{.007} & \result{.579}{.007} &
    \result{.596}{.007} & \result{.554}{.006} & \result{.585}{.007} & \result{.549}{.006} \\
         
    \rowcolor{blue!10}
    & \ours &
        \bestresult{.608}{.007} & \bestresult{.588}{.006} & \bestresult{.587}{.006} & \bestresult{.602}{.007} &
        \bestresult{.620}{.007} & \bestresult{.597}{.006} & \bestresult{.610}{.007} & \bestresult{.568}{.007} \\
    \midrule
    \multirow{5}{*}{GPT-4.1-mini} &
      Direct &
        \result{.582}{.006} & \result{.534}{.006} & \result{.576}{.006} & \result{.579}{.007} &
        \result{.589}{.007} & \result{.531}{.006} & \result{.584}{.007} & \result{.543}{.006} \\
    & CoT &
        \result{.581}{.007} & \result{.525}{.006} & \result{.568}{.006} & \result{.571}{.007} &
        \result{.580}{.007} & \result{.523}{.006} & \result{.575}{.007} & \result{.538}{.006} \\
    & Self-Refine &
    \result{.608}{.007} & \result{.544}{.007} & \result{.571}{.007} & \result{.593}{.008} &
    \result{.610}{.007} & \result{.553}{.007} & \result{.580}{.008} & \result{.566}{.006} \\
    & MAD &
    \result{.598}{.007} & \result{.553}{.006} & \result{.569}{.006} & \result{.592}{.008} &
    \result{.616}{.006} & \result{.559}{.007} & \result{.586}{.008} & \result{.570}{.006} \\
    \rowcolor{blue!10}
    & \ours &
        \bestresult{.614}{.007} & \bestresult{.594}{.006} & \bestresult{.585}{.006} & \bestresult{.608}{.007} &
        \bestresult{.623}{.007} & \bestresult{.601}{.006} & \bestresult{.604}{.007} & \bestresult{.587}{.006} \\
    \midrule
    \multirow{5}{*}{GPT-4.1} &
      Direct &
        \result{.578}{.006} & \result{.531}{.006} & \result{.573}{.006} & \result{.575}{.007} &
        \result{.584}{.007} & \result{.530}{.006} & \result{.577}{.006} & \result{.541}{.006} \\
    & CoT &
        \result{.582}{.006} & \result{.537}{.006} & \result{.572}{.006} & \result{.577}{.006} &
        \result{.585}{.007} & \result{.533}{.006} & \result{.578}{.007} & \result{.539}{.006} \\
    & Self-Refine &
    \result{.595}{.007} & \result{.556}{.006} & \result{.574}{.006} & \result{.594}{.007} &
    \result{.605}{.008} & \result{.543}{.006} & \result{.575}{.007} & \result{.557}{.007} \\
    & MAD &
    \result{.599}{.007} & \result{.561}{.007} & \result{.576}{.007} & \result{.594}{.007} &
    \result{.609}{.010} & \result{.563}{.006} & \result{.580}{.008} & \result{.567}{.006} \\
    \rowcolor{blue!10}
    & \ours &
        \bestresult{.621}{.007} & \bestresult{.622}{.007} & \bestresult{.582}{.006} & \bestresult{.619}{.007} &
        \bestresult{.640}{.007} & \bestresult{.634}{.006} & \bestresult{.602}{.007} & \bestresult{.603}{.007} \\
    \bottomrule
  \end{tabular}
  \caption{%
    Performance comparison between \ours and baselines on DeliberationBank task. We present the results on the two sub-categories of the questions: OpenQA and BinaryQA. Values are mean $\pm$ 95\% CI. We report scores for representativeness (Represent.), informativeness (Inform.), neutrality (Neutral.), and policy approval (Policy; higher is better). Best scores per column are in \textbf{bold}.
  }
  \label{tab:deliberation}
\end{table*}

\paragraph{Experiment Details}

We primarily evaluate teams of four participants. For each question, we cluster the crowd-sourced comments into four groups and sample one comment from each cluster, forming a team intended to cover qualitatively different stances. We sample a total of 500 team configurations. 
We follow DeliberationBank protocol~\citep{zhu2025can} to score the outputs, and in addition using LLM as a judge~\citep{li2025generation} to perform pairwise comparison.

\paragraph{Baselines and Metrics}
We compare with: 1. \textbf{Direct summary} that prompts an LLM to summarize the comments, 2. \textbf{Chain-of-Thought} (CoT) that prompts LLM to think before generating a final summary~\citep{wei2022chain}, 3. \textbf{Self-Refinement} (Self-Refine), iteratively refining summaries without structured interaction~\cite{madaan2023self}, 4. \textbf{Multi-Agent Debate} (MAD), using generic agents conducting four rounds of debate~\citep{du2023improving}.
We report four dimensions from DeliberationBank: representativeness, informativeness, neutrality, and policy approval. A detailed description of the four metrics and their significance to open-ended decision making is presented in Table~\ref{tab:task2_metrics_decision_support}.

\subsection{Experiment Results}


\paragraph{TeamFusion consistently outperforms baselines.}
Table~\ref{tab:deliberation} shows consistent gains from \ours across base models and question types. We focus on \textbf{representativeness} as the primary metric because it directly captures our goal of preserving diverse viewpoints. \ours yields the largest improvements on representativeness, and these gains co-occur with strong increases in informativeness and policy approval, suggesting that the additional structured discussion surfaces missing considerations that make summaries more decision-ready. Importantly, neutrality remains comparable to baselines, indicating that improved viewpoint coverage does not come from introducing more polarized or editorial language.

\begin{table}[t]
  \centering
  \small
  \setlength{\tabcolsep}{3pt}
  \begin{tabular}{l|cccc}
    \toprule
    \addlinespace[1pt]
    \midrule
    & \multicolumn{4}{c}{Win / Tie / Loss (\%)} \\
    \cmidrule(lr){2-5}
    Model & Represent. & Inform. & Neutral. & Policy \\
    \midrule
    Llama-70B &
      71 / 28 / 1 &
      95 / 0 / 5 &
      51 / 43 / 6 &
      97 / 0 / 3 \\
      GPT-4.1-mini &
      72 / 26 / 2 &
      96 / 0 / 4 &
      27 / 61 / 12 &
      96 / 0 / 4 \\
    GPT-4.1 &
      93 / 7 / 0 &
      98 / 0 / 2 &
      46 / 49 / 5 &
      99 / 0 / 1 \\
    \bottomrule
  \end{tabular}
  \caption{
    Win/Tie/Loss rate of \ours outcome against direct summary across four metrics. Higher win rates indicate stronger relative performance.
  }
  \label{tab:winrates}
\end{table}

To complement these aggregate scores, we also conduct a pairwise comparison between \ours-generated summaries and direct summaries using an LLM-as-a-judge. We randomly sample 300 \ours outcomes (100 for each base model), pair them with the corresponding direct summaries, and prompt GPT-4.1-mini to decide which summary is better. To avoid any positional bias of the LLM judge~\citep{shi2024judging}, we randomized the order of the summaries in the prompt. As shown in Table~\ref{tab:winrates}, \ours wins overwhelmingly on informativeness and policy approval, and wins on representativeness in the large majority of cases.

\paragraph{Performance gains stem from personalized interaction.} We compare TeamFusion against compute-matched Self-Refine and MAD baselines. While MAD involves structured discussion among agents, its generic approach without personalization leads to limited viewpoint diversity and narrower coverage. Similarly, self-refinement provides an iterative reasoning structure but lacks interactive discussion. In contrast, \ours's performance boost is primarily attributable to its personalized agent interactions, highlighting the critical role of personalization and structured discussion. This analysis confirms that the observed improvements are not merely due to increased computational budget but rather due to the strategic combination of personalized representation and iterative debating.


\begin{table}[t]
  \centering
  \small
  \setlength{\tabcolsep}{5pt}
  \begin{tabular}{l*{4}{c}}
    \toprule
    \addlinespace[1pt]
    \midrule
    Method & Represent. & Inform. & Neutral. & Policy \\
    \midrule

    \rowcolor{gray!15}
    \multicolumn{5}{l}{\textbf{Team size: 6}} \\
    Direct &
        \result{.582}{.010} &
        \result{.537}{.008} &
        \result{.575}{.009} &
        \result{.560}{.010} \\
    CoT &
        \result{.576}{.010} &
        \result{.529}{.008} &
        \result{.569}{.009} &
        \result{.556}{.010} \\
    Self-Refine &
        \result{.607}{.010} &
        \result{.561}{.009} &
        \result{.576}{.009} &
        \result{.581}{.010} \\
    MAD &
        \result{.600}{.009} &
        \result{.573}{.009} &
        \result{.579}{.009} &
        \result{.587}{.010} \\
    \rowcolor{blue!10}
    \ours &
        \bestresult{.622}{.010} &
        \bestresult{.617}{.008} &
        \bestresult{.593}{.010} &
        \bestresult{.608}{.010} \\
    \midrule

    \rowcolor{gray!15}
    \multicolumn{5}{l}{\textbf{Team size: 8}} \\
    Direct &
        \result{.580}{.009} &
        \result{.540}{.008} &
        \result{.580}{.008} &
        \result{.556}{.009} \\
    CoT &
        \result{.575}{.009} &
        \result{.530}{.008} &
        \result{.573}{.009} &
        \result{.550}{.009} \\
    Self-Refine &
        \result{.603}{.009} &
        \result{.568}{.008} &
        \result{.578}{.009} &
        \result{.576}{.010} \\
    MAD &
        \result{.605}{.009} &
        \result{.570}{.008} &
        \result{.580}{.009} &
        \result{.578}{.010} \\
    \rowcolor{blue!10}
    \ours &
        \bestresult{.621}{.009} &
        \bestresult{.627}{.007} &
        \bestresult{.596}{.009} &
        \bestresult{.609}{.009} \\
    \midrule

    \rowcolor{gray!15}
    \multicolumn{5}{l}{\textbf{Team size: 10}} \\
    Direct &
        \result{.578}{.007} &
        \result{.546}{.007} &
        \result{.579}{.008} &
        \result{.554}{.008} \\
    CoT &
        \result{.574}{.008} &
        \result{.535}{.007} &
        \result{.571}{.007} &
        \result{.550}{.008} \\
    Self-Refine &
        \result{.604}{.009} &
        \result{.574}{.008} &
        \result{.578}{.009} &
        \result{.576}{.010} \\
    MAD &
        \result{.602}{.009} &
        \result{.572}{.007} &
        \result{.580}{.008} &
        \result{.579}{.009} \\
    \rowcolor{blue!10}
    \ours &
        \bestresult{.619}{.008} &
        \bestresult{.637}{.008} &
        \bestresult{.596}{.009} &
        \bestresult{.608}{.009} \\
    \bottomrule
  \end{tabular}
  \caption{
    Performance comparison between \ours and baselines on the DeliberationBank task
    for different team sizes. Values are mean $\pm$ 95\% CI. Best scores per column are in \textbf{bold}.
  }
  \label{tab:size}
\end{table}

\paragraph{\ours demonstrates gains across different team sizes.}
We then investigate the effectiveness of team size. We fix the total number of sampled team configurations to 100 and use GPT-4.1-mini as the backbone. Results are shown in Table~\ref{tab:size}. Across different team sizes, \ours consistently outperforms baselines on the key representativeness metric. \ours can improve representativeness by about 0.04 absolute gain over the baselines, with non-overlapping confidence intervals. This demonstrates the scalability and generalization capability of \ours.

\paragraph{Iterative refinement brings gains.}
We fix the total number of sampled team configurations to 100 and run \ours over up to three iterations (i.e., two refinements). The results of iterative refinement are shown in Table~\ref{tab:refinement}. For all models, adding iterative refinement brings gains to the representativeness and informativeness of the final summary across the iterations. Neutrality and policy alignment also improve, though with smaller margins, suggesting that additional rounds are particularly effective at surfacing missing considerations rather than merely smoothing tone. 

\begin{table}[t]
  \centering
  \small
  \setlength{\tabcolsep}{5pt}  
  \begin{tabular}{l*{4}{c}}
    \toprule
    \addlinespace[1pt]
    \midrule
     & Represent. & Inform. & Neutral. & Policy \\
    \midrule

    \rowcolor{gray!15}
    \multicolumn{5}{l}{\textbf{Model: Llama-3.3-70B}} \\
    Base &
        \result{.582}{.007} &
        \result{.541}{.006} &
        \result{.568}{.006} &
        \result{.552}{.006} \\
    + Iter 1&
        \result{.601}{.006} &
        \result{.563}{.005} &
        \result{.579}{.006} &
        \result{.566}{.006} \\
    + Iter 2 &
        \result{\mathbf{.618}}{.006} &
        \result{\mathbf{.581}}{.005} &
        \result{\mathbf{.592}}{.006} &
        \result{\mathbf{.581}}{.006} \\
    \midrule

    \rowcolor{gray!15}
    \multicolumn{5}{l}{\textbf{Model: GPT-4.1-mini}} \\
    Base &
        \result{.622}{.018} &
        \result{.596}{.014} &
        \result{.596}{.016} &
        \result{.599}{.016} \\
    + Iter 1 &
        \result{.633}{.018} &
        \result{.625}{.015} &
        \result{\mathbf{.604}}{.014} &
        \result{.618}{.018} \\
    + Iter 2 &
        \result{\mathbf{.637}}{.018} &
        \result{\mathbf{.638}}{.015} &
        \result{.599}{.015} &
        \result{\mathbf{.619}}{.016} \\
    \midrule

    \rowcolor{gray!15}
    \multicolumn{5}{l}{\textbf{Model: GPT-4.1}} \\
    Base &
        \result{.630}{.018} &
        \result{.615}{.014} &
        \result{.591}{.016} &
        \result{.606}{.016} \\
    + Iter 1 &
        \result{.646}{.019} &
        \result{.655}{.014} &
        \result{.599}{.015} &
        \result{.635}{.017} \\
    + Iter 2 &
        \result{\mathbf{.655}}{.017} &
        \result{\mathbf{.686}}{.014} &
        \result{\mathbf{.603}}{.015} &
        \result{\mathbf{.643}}{.018} \\
    \bottomrule

  \end{tabular}
  \caption{
    Performance of different models under iterative refinement. 
    Values are mean $\pm$ 95\% CI.
  }
  \label{tab:refinement}
\end{table}

\section{Task 2: Visual Design}


We then study \ours in a \emph{human-centric, multi-modal} workflow grounded in \emph{real industry practice}.


\subsection{Problem Motivation}
Creative alignment is a significant pain point in professional design. Unlike close-ended tasks with objective ``gold labels,'' design briefs are open-ended and subject to interpretation. This ambiguity introduces friction in industry practice: teams must expend significant effort negotiating trade-offs between aesthetics, brand tone, and constraints. 

We motivate this task by empirically quantifying this friction. In our preliminary analysis of professional designers' preferences (detailed in Sec 5.4), we observed that experts given the exact same brief and assets exhibited remarkably low agreement on quality. In 70\% of cases, agreement was indistinguishable from random chance. This validates that divergent interpretation is a natural and pervasive bottleneck.

\subsection{Task Setup}
Each scenario consists of a client brief clarifying the requirements for an advertisement design and a set of candidate ad thumbnails. A team of professional designers rank the candidates and provide justifications. 
\ours runs proxy agent discussion and produces remixed design images intended to better align with designers' expressed constraints while making the underlying points of agreement and disagreement actionable for downstream selection. We evaluate whether \ours generated images can replace original team's favorites.

\subsection{Experiment Protocol}

We introduce a human-in-the-loop protocol to evaluate \ours. Unlike static benchmark evaluations, this protocol allows us to scale realistic team interactions while keeping professional designers as the ultimate ground truth for decision quality.

\paragraph{Phase 0: Realistic Scenario Construction}
We construct 50 high-quality design scenarios derived from real social media advertising campaigns~\cite{yamaguchi2021canvasvae}. Each scenario includes a professional client brief and a set of diverse candidate designs. All scenarios have been validated by two external senior designers for realism. Full construction details can be found in Appendix~\ref{sec:data}.

\begin{figure}[t]
  \centering
  \includegraphics[width=\columnwidth]{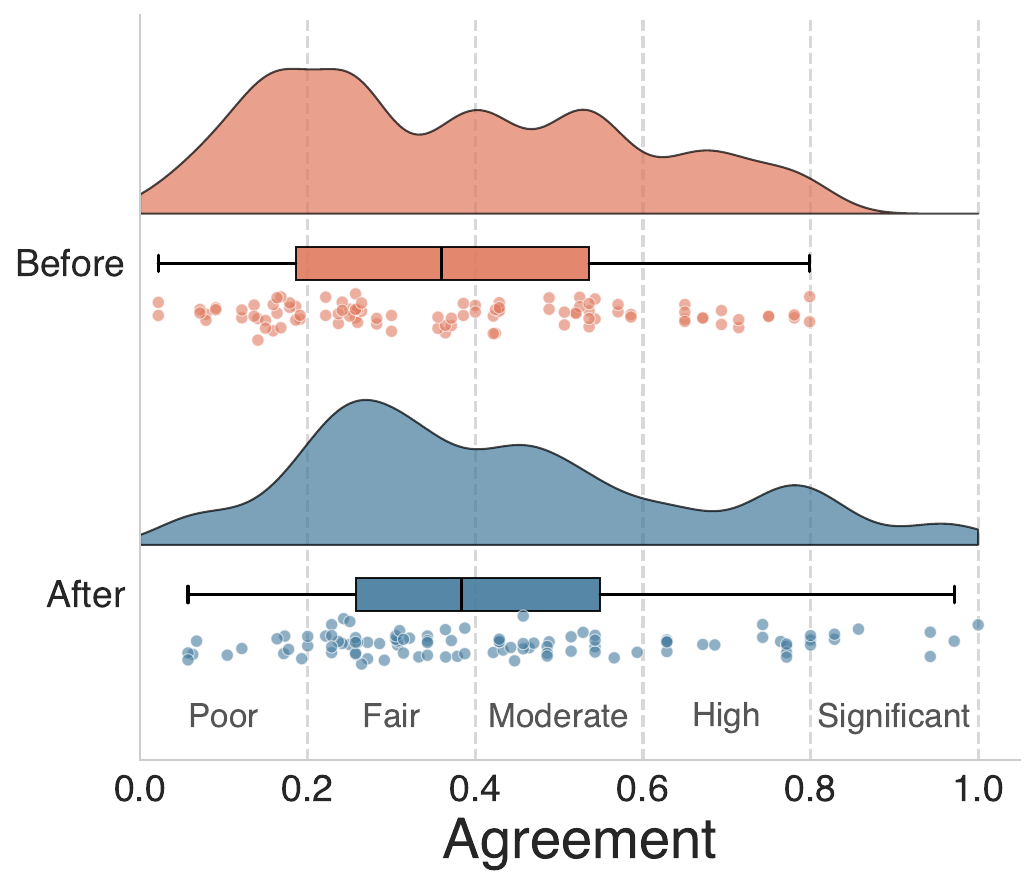}
  \caption{The distribution of agreement scores to measure dataset-wide agreement before and after \ours's execution. The data is categorized into five value ranges to interpret agreement strength. Agreements across 100 team settings after running \ours show a dataset-wide move towards higher agreement.}
  \label{fig:w_dist}
\end{figure}

\paragraph{Phase 1: Preference collection}
We recruit 9 professional designers to annotate scenarios asynchronously. For each assigned scenario, designer ranks the six options and writes a brief justification. Each scenario received at least four independent annotations, serving as the ``seed'' evidence that conditions our proxy agents.

\paragraph{Phase 2: Simulation with nominal teams}
For each scenario, we form two nominal team settings from asynchronous annotations: \textbf{Full-Team} (all available annotations) and \textbf{Small-Team} (a random subset of two designers). We run \ours for three iterations, producing one new remixed design per iteration.This yields 100 \ours runs and 300 remixed design candidates. We also record each proxy agent's discussion comments about the generated candidates for later analysis.

\paragraph{Phase 3: Designer re-evaluation}

To determine if the system successfully facilitated convergence, we close the loop by returning the generated outputs to the original human designers. We combine the team's initial top three options via Borda count with the three \ours-generated options. Designers then re-rank the combined set and rate whether their proxy agent's commentary aligns with their own reasoning.


\subsection{Experiment Results}
\label{sec:results}

Our analysis reveals three main findings. First, we empirically verify the motivating problem of divergent preferences in teams. Second, we show that \ours can generate consensus-oriented remixes that successfully induce convergence. Finally, we show that real designers largely agree with the debate commentary made by their delegate agents, indicating that simulated debates are well-grounded.

\paragraph{Finding 1: Divergent interpretations are a real, salient problem.}
As outlined in our motivation, we hypothesized that professional designers hold conflicting interpretations of the same brief. Our analysis of the pre-discussion ranking data confirms this friction is substantial. To quantify this, we calculate Kendall's Coefficient of Concordance ($W$) on the independent rankings provided in Phase 1.
As shown in Figure~\ref{fig:w_dist} (top), the agreement among designers is consistently low, with a mean of 0.37 (falling into the ``Fair Agreement'' range). Notably, 84\% of scenarios fall into the "Moderate" or lower agreement categories, and in 70\% of cases, the agreement among professionals is not statistically significant ($p \ge 0.05$). This widespread lack of consensus in the real data confirms that our motivating problem is natural-arising and salient, providing strong empirical evidence in support of systems like \ours.

\paragraph{Finding 2: \ours can support team convergence by generating consensus-oriented designs.}
Our results reveal two ways in which \ours-generated design revisions meaningfully modify the output of creative teams.

\begin{figure}[t]
  \centering
  \includegraphics[width=0.85\columnwidth]{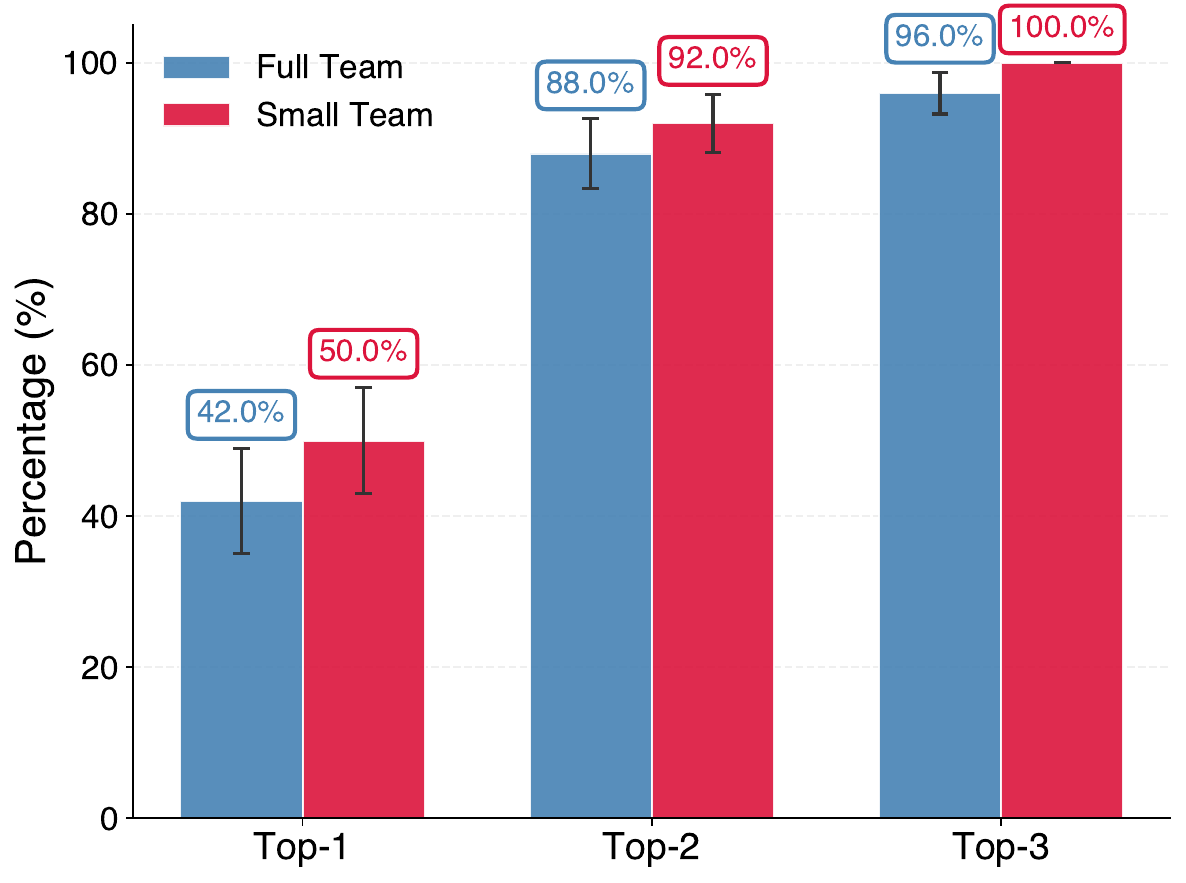}
  \caption{The rate of \ours-generated images appearing in the final top-ranked selections. Error bars represent the 95\% confidence interval.}
  \label{fig:any}
\end{figure}

The results presented in Figure~\ref{fig:any} show that \textbf{\ours-generated options become the single top-ranked option across the team in nearly half of all test cases}, displacing the original team-wide favorites. Notably, this indicates that the execution of \ours can be seen as a generative AI feature with significant team-wide acceptance rate under the strictest decision-making scenario, that is, the team decides to move forward with the single best design only.

Under less strict decision-making scenarios, \ours also shows potential for contributing to teams' outputs. We find that \ours-generated option appeared in the top-two rankings in a remarkable 88\% of Full-Team and 92\% of Small-Team test cases. This finding is noteworthy because, in the creative decision-making space, teams may use not only the single best option out of group ideation, but actually a short-list of top options; for example, teams may steer the creative process by revising on one of them.

More globally, Figure~\ref{fig:w_dist} shows how there is a dataset-wide move towards higher agreement after the exposure to \ours outputs: mean Kendall's $W$ rises from 0.37 to 0.43. Qualitatively, the dataset mean moves from only ``Fair Agreement'' before to ``Moderate Agreement'' after \ours, which further indicates the effectiveness of our system in supporting convergence. 

\begin{figure}[t]
  \centering
  \includegraphics[width=\columnwidth]{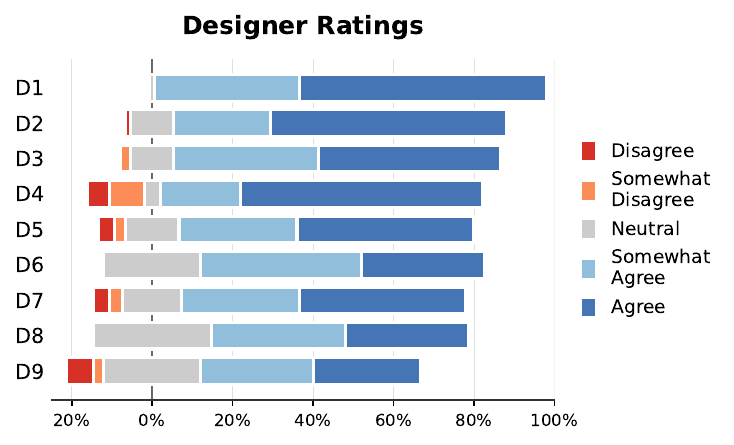}
  \caption{Distribution of annotator ratings for agreement with agent-generated commentary, grouped by designers. The results show an overwhelmingly positive perception.}
  \label{fig:likert}
\end{figure}

\paragraph{Finding 3: Designers feel largely represented by their proxy agents.} 
As shown in Figure~\ref{fig:likert}, the results were \textbf{overwhelmingly positive}. The scores were strongly skewed toward agreement, with a mean of 4.06 ($\sigma = 1.07$). Over 75\% of all comments received a positive score. This indicates that designers broadly perceived the outputs of their proxy agents as natural-sounding and representative of their own design rationales.
This positive perception was highly consistent across our participants. An analysis of designer-level means showed a narrow range, with the lowest average rating being 3.53. This indicates that even the most critical participant found the commentary to be representative. The low standard deviation of these designer means ($\sigma=0.29$) further reinforces that this high level of agreement is a shared, consistent finding.

\subsection{Live User Study Results}
To complement our asynchronous evaluation, we run a live, controlled within-team study to test whether \ours facilitates convergence in an end-to-end workflow where participants create and revise designs. We recruit six participants and formed two teams of three, with each team completing two tasks in a counterbalanced crossover design\footnote{Full details can be found in Appendix~\ref{app:live-study}}. Due to the small number of participants, we report this live study as a pilot with descriptive results rather than a statistically powered evaluation. As shown in Table~\ref{tab:live-study}, using \ours leads to faster team decisions compared to free-form discussion, while also improving participants’ perceived representativeness, clarity of trade-offs, and overall satisfaction with the team outcome. After experiencing both workflows, a strong majority of participants explicitly preferred \ours over free-form discussion, showcasing the effectiveness over unconstrained collaboration.

\begin{table}[t]
\centering
\small
\resizebox{\columnwidth}{!}{%
\begin{tabular}{lcccc}
\toprule
\textbf{Metric} & \textbf{Discussion} & \textbf{\ours} \\
\midrule
Decision Time (min) $\downarrow$ & 18.0 & \textbf{12.4}  \\
\midrule
Q1: Representative $\uparrow$ & 3.7 & \textbf{4.3}  \\
Q2: Clarity $\uparrow$ & 3.5 & \textbf{3.8} \\
Q3: Satisfaction $\uparrow$ & 3.5 & \textbf{4.2}  \\
\midrule
Preferred $\uparrow$ & 1/6 & \textbf{5/6}  \\
\bottomrule
\end{tabular}%
}
\caption{Live within-team study results. Each team completed two briefs in a counterbalanced crossover design.}
\label{tab:live-study}
\end{table}

\section{Case Study}
We present a case study on the effectiveness in Figure~\ref{fig:case_study}. Due to page limit, we defer the detailed analysis in Appendix~\ref{case_study}. The core takeaway is that \ours can better preserve fine-grained, participant-specific content than direct aggregation.
We also present a case study on the failure mode of \ours in Appendix~\ref{failure_mode}.

\section{Conclusion}
Open-ended team decisions require deliverables that make trade-offs and disagreements visible rather than averaging them away, yet common aggregation methods often erase minority or conditional viewpoints and reduce auditability. We addressed this challenge with \ours, a multi-agent framework that shifts the paradigm from direct aggregation to modeled interaction. By representing participants with preference-grounded proxy agents and orchestrating structured debates, TeamFusion fully develops the consensus-seeking process to produce editable, rationale-backed deliverables. Evaluations on two teamwork tasks show that explicitly modeling interaction improves viewpoint coverage and decision usefulness and can induce greater convergence, validating the potential of AI to facilitate human collaboration.

\section*{Limitations}

Our findings show that \ours can successfully support the creative convergence process, producing consensus-inducing design revisions grounded in natural-sounding, agreeable rationales. However, our work has limitations that shed light on important directions for future research. The current implementation assumes a flat hierarchy in the team, not accounting for different roles (e.g., art directors managing the team, or even clients themselves in the loop) and seniority levels (e.g., senior vs. junior designers). Even more realistic professional settings may warrant slightly different assumptions, with implications to our current modeling decisions.

\section*{Ethics Statement}

\ours is designed to \textbf{support}---and \textbf{not} substitute---creative teams in their decision-making, keeping human taste and judgment front and center in how the system is used and guided. The configuration of iterations and the production of iterative outputs further aligns with the principle that human teams are not only the users guiding the system but also the ultimate decision-makers determining whether and to what extent \ours's outputs are useful. For the same reason, we propose that the underlying prompts producing the outputs and their provenance back to the multi-agent debates are kept transparent to users, such that they can be reviewed and remixed themselves, providing full human-in-the-loop control.


\bibliography{ref}

\begin{thebibliography}{66}
\providecommand{\natexlab}[1]{#1}

\bibitem[{Acar and Runco(2019)}]{acar2019divergent}
Selcuk Acar and Mark~A Runco. 2019.
\newblock Divergent thinking: New methods, recent research, and extended theory.
\newblock \emph{Psychology of aesthetics, creativity, and the arts}, 13(2):153.

\bibitem[{Agarwal et~al.(2024)Agarwal, Fabbri, Risher, Laban, Joty, and Wu}]{agarwal-etal-2024-prompt}
Divyansh Agarwal, Alexander Fabbri, Ben Risher, Philippe Laban, Shafiq Joty, and Chien-Sheng Wu. 2024.
\newblock \href {https://doi.org/10.18653/v1/2024.emnlp-industry.94} {Prompt leakage effect and mitigation strategies for multi-turn {LLM} applications}.
\newblock In \emph{Proceedings of the 2024 Conference on Empirical Methods in Natural Language Processing: Industry Track}, pages 1255--1275, Miami, Florida, US. Association for Computational Linguistics.

\bibitem[{Bhaskar et~al.(2023)Bhaskar, Fabbri, and Durrett}]{bhaskar-etal-2023-prompted}
Adithya Bhaskar, Alex Fabbri, and Greg Durrett. 2023.
\newblock \href {https://doi.org/10.18653/v1/2023.findings-acl.591} {Prompted opinion summarization with {GPT}-3.5}.
\newblock In \emph{Findings of the Association for Computational Linguistics: ACL 2023}, pages 9282--9300, Toronto, Canada. Association for Computational Linguistics.

\bibitem[{Black(1948)}]{black1948rationale}
Duncan Black. 1948.
\newblock On the rationale of group decision-making.
\newblock \emph{Journal of political economy}, 56(1):23--34.

\bibitem[{Carletta et~al.(2006)Carletta, Ashby, Bourban, Flynn, Guillemot, Hain, Kadlec, Karaiskos, Kraaij, Kronenthal, Lathoud, Lincoln, Lisowska, McCowan, Post, Reidsma, and Wellner}]{10.1007/11677482_3}
Jean Carletta, Simone Ashby, Sebastien Bourban, Mike Flynn, Mael Guillemot, Thomas Hain, Jaroslav Kadlec, Vasilis Karaiskos, Wessel Kraaij, Melissa Kronenthal, Guillaume Lathoud, Mike Lincoln, Agnes Lisowska, Iain McCowan, Wilfried Post, Dennis Reidsma, and Pierre Wellner. 2006.
\newblock The ami meeting corpus: A pre-announcement.
\newblock In \emph{Machine Learning for Multimodal Interaction}, pages 28--39, Berlin, Heidelberg. Springer Berlin Heidelberg.

\bibitem[{Chan et~al.(2023)Chan, Chen, Su, Yu, Xue, Zhang, Fu, and Liu}]{chan2023chateval}
Chi-Min Chan, Weize Chen, Yusheng Su, Jianxuan Yu, Wei Xue, Shanghang Zhang, Jie Fu, and Zhiyuan Liu. 2023.
\newblock Chateval: Towards better llm-based evaluators through multi-agent debate.
\newblock \emph{arXiv preprint arXiv:2308.07201}.

\bibitem[{Chase(2022)}]{chase_langchain_2022}
Harrison Chase. 2022.
\newblock Langchain.
\newblock \url{https://github.com/langchain-ai/langchain}.

\bibitem[{Chen et~al.(2024)Chen, Lu, Du, Rejtig, Bagley, Horn, and Wilensky}]{Chen2024LearningAMA}
John Chen, Xi~Lu, Yuzhou Du, Michael Rejtig, Ruth Bagley, Mike Horn, and Uri Wilensky. 2024.
\newblock \href {https://api.semanticscholar.org/CorpusId:267320240} {Learning agent-based modeling with llm companions: Experiences of novices and experts using chatgpt \& netlogo chat}.
\newblock \emph{Proceedings of the 2024 CHI Conference on Human Factors in Computing Systems}.

\bibitem[{Chern et~al.(2024)Chern, Fan, and Liu}]{chern2024combating}
Steffi Chern, Zhen Fan, and Andy Liu. 2024.
\newblock Combating adversarial attacks with multi-agent debate.
\newblock \emph{arXiv preprint arXiv:2401.05998}.

\bibitem[{Chiang et~al.(2024)Chiang, Lu, Li, and Yin}]{chiang2024enhancing}
Chun-Wei Chiang, Zhuoran Lu, Zhuoyan Li, and Ming Yin. 2024.
\newblock Enhancing ai-assisted group decision making through llm-powered devil's advocate.
\newblock In \emph{Proceedings of the 29th International Conference on Intelligent User Interfaces}, pages 103--119.

\bibitem[{del Cerro et~al.(1998)del Cerro, Herzig, Longin, and Rifi}]{del1998belief}
Luis~Fari{\~n}as del Cerro, Andreas Herzig, Dominique Longin, and Omar Rifi. 1998.
\newblock Belief reconstruction in cooperative dialogues.
\newblock In \emph{International Conference on Artificial Intelligence: Methodology, Systems, and Applications}, pages 254--266. Springer.

\bibitem[{Dowling and St.~Louis(2000)}]{10.1016/S0167-9236(00)00073-7}
Karen~L. Dowling and Robert~D. St.~Louis. 2000.
\newblock \href {https://doi.org/10.1016/S0167-9236(00)00073-7} {Asynchronous implementation of the nominal group technique: is it effective?}
\newblock \emph{Decis. Support Syst.}, 29(3):229–248.

\bibitem[{Du et~al.(2023)Du, Li, Torralba, Tenenbaum, and Mordatch}]{du2023improving}
Yilun Du, Shuang Li, Antonio Torralba, Joshua~B Tenenbaum, and Igor Mordatch. 2023.
\newblock Improving factuality and reasoning in language models through multiagent debate.
\newblock In \emph{Forty-first International Conference on Machine Learning}.

\bibitem[{Fisher(1970)}]{fisher1970decision}
B~Aubrey Fisher. 1970.
\newblock Decision emergence: Phases in group decision-making.
\newblock \emph{Communications Monographs}, 37(1):53--66.

\bibitem[{Govers et~al.(2024)Govers, Velloso, Kostakos, and Goncalves}]{10.1145/3613904.3642322}
Jarod Govers, Eduardo Velloso, Vassilis Kostakos, and Jorge Goncalves. 2024.
\newblock \href {https://doi.org/10.1145/3613904.3642322} {Ai-driven mediation strategies for audience depolarisation in online debates}.
\newblock In \emph{Proceedings of the 2024 CHI Conference on Human Factors in Computing Systems}, CHI '24, New York, NY, USA. Association for Computing Machinery.

\bibitem[{Greshake et~al.(2023)Greshake, Abdelnabi, Mishra, Endres, Holz, and Fritz}]{greshake2023not}
Kai Greshake, Sahar Abdelnabi, Shailesh Mishra, Christoph Endres, Thorsten Holz, and Mario Fritz. 2023.
\newblock Not what you've signed up for: Compromising real-world llm-integrated applications with indirect prompt injection.
\newblock In \emph{Proceedings of the 16th ACM workshop on artificial intelligence and security}, pages 79--90.

\bibitem[{Guo et~al.(2024)Guo, Huang, Liu, Fan, V{\'e}lez, Wu, Wang, Griffiths, and Wang}]{guo2024embodied}
Xudong Guo, Kaixuan Huang, Jiale Liu, Wenhui Fan, Natalia V{\'e}lez, Qingyun Wu, Huazheng Wang, Thomas~L Griffiths, and Mengdi Wang. 2024.
\newblock Embodied llm agents learn to cooperate in organized teams.
\newblock \emph{arXiv preprint arXiv:2403.12482}.

\bibitem[{Han et~al.(2024)Han, Zhang, Jin, and Xu}]{han2024llm}
Shanshan Han, Qifan Zhang, Weizhao Jin, and Zhaozhuo Xu. 2024.
\newblock Llm multi-agent systems: Challenges and open problems.
\newblock \emph{arXiv preprint arXiv:2402.03578}.

\bibitem[{Hong et~al.(2023)Hong, Zhuge, Chen, Zheng, Cheng, Wang, Zhang, Wang, Yau, Lin et~al.}]{hong2023metagpt}
Sirui Hong, Mingchen Zhuge, Jonathan Chen, Xiawu Zheng, Yuheng Cheng, Jinlin Wang, Ceyao Zhang, Zili Wang, Steven Ka~Shing Yau, Zijuan Lin, and 1 others. 2023.
\newblock Metagpt: Meta programming for a multi-agent collaborative framework.
\newblock In \emph{The Twelfth International Conference on Learning Representations}.

\bibitem[{Hu et~al.(2025)Hu, Chan, Li, and Yin}]{hu2025debate}
Zhe Hu, Hou~Pong Chan, Jing Li, and Yu~Yin. 2025.
\newblock Debate-to-write: A persona-driven multi-agent framework for diverse argument generation.
\newblock In \emph{Proceedings of the 31st International Conference on Computational Linguistics}, pages 4689--4703.

\bibitem[{Huang et~al.(2025)Huang, Yu, Ma, Zhong, Feng, Wang, Chen, Peng, Feng, Qin et~al.}]{huang2025survey}
Lei Huang, Weijiang Yu, Weitao Ma, Weihong Zhong, Zhangyin Feng, Haotian Wang, Qianglong Chen, Weihua Peng, Xiaocheng Feng, Bing Qin, and 1 others. 2025.
\newblock A survey on hallucination in large language models: Principles, taxonomy, challenges, and open questions.
\newblock \emph{ACM Transactions on Information Systems}, 43(2):1--55.

\bibitem[{Huang et~al.(2023)Huang, Tian, Fayek, and Zhang}]{huang-etal-2023-examining}
Nannan Huang, Lin Tian, Haytham Fayek, and Xiuzhen Zhang. 2023.
\newblock \href {https://doi.org/10.18653/v1/2023.wassa-1.14} {Examining bias in opinion summarisation through the perspective of opinion diversity}.
\newblock In \emph{Proceedings of the 13th Workshop on Computational Approaches to Subjectivity, Sentiment, {\&} Social Media Analysis}, pages 149--161, Toronto, Canada. Association for Computational Linguistics.

\bibitem[{Huang et~al.(2024)Huang, Liu, Ko, Wu, Wang, Zhang, and Tang}]{Huang2024SelectivePTA}
Qiushi Huang, Xubo Liu, Tom Ko, Boyong Wu, Wenwu Wang, Yu~Zhang, and Lilian Tang. 2024.
\newblock \href {https://api.semanticscholar.org/CorpusId:270737891} {Selective prompting tuning for personalized conversations with llms}.
\newblock \emph{ArXiv}, abs/2406.18187.

\bibitem[{Johansson(2002)}]{johansson2002user}
Pontus Johansson. 2002.
\newblock User modeling in dialog systems.
\newblock \emph{St. Anna Report SAR}, pages 02--2.

\bibitem[{Kiesler and Sproull(1992)}]{kiesler1992group}
Sara Kiesler and Lee Sproull. 1992.
\newblock Group decision making and communication technology.
\newblock \emph{Organizational behavior and human decision processes}, 52(1):96--123.

\bibitem[{Kim et~al.(2024)Kim, Park, Jeong, Chan, Xu, McDuff, Lee, Ghassemi, Breazeal, and Park}]{kim2024mdagents}
Yubin Kim, Chanwoo Park, Hyewon Jeong, Yik~S Chan, Xuhai Xu, Daniel McDuff, Hyeonhoon Lee, Marzyeh Ghassemi, Cynthia Breazeal, and Hae~W Park. 2024.
\newblock Mdagents: An adaptive collaboration of llms for medical decision-making.
\newblock \emph{Advances in Neural Information Processing Systems}, 37:79410--79452.

\bibitem[{Kobsa(1989)}]{kobsa1989taxonomy}
Alfred Kobsa. 1989.
\newblock A taxonomy of beliefs and goals for user models in dialog systems.
\newblock In \emph{User models in dialog systems}, pages 52--68. Springer.

\bibitem[{Kraemer and King(1988)}]{kraemer1988computer}
Kenneth~L Kraemer and John~Leslie King. 1988.
\newblock Computer-based systems for cooperative work and group decision making.
\newblock \emph{ACM Computing Surveys (CSUR)}, 20(2):115--146.

\bibitem[{Kwon et~al.(2023)Kwon, Lee, Kim, Lee, Kim, and Davis}]{Kwon2023WhatWAA}
D.~Kwon, Sunwoo Lee, Ki~Hyun Kim, Seojin Lee, Tae-Yoon Kim, and Eric Davis. 2023.
\newblock \href {https://api.semanticscholar.org/CorpusId:259089085} {What, when, and how to ground: Designing user persona-aware conversational agents for engaging dialogue}.
\newblock In \emph{Annual Meeting of the Association for Computational Linguistics}.

\bibitem[{Laban et~al.(2023)Laban, Kry{\'s}ci{\'n}ski, Agarwal, Fabbri, Xiong, Joty, and Wu}]{laban2023summedits}
Philippe Laban, Wojciech Kry{\'s}ci{\'n}ski, Divyansh Agarwal, Alexander~Richard Fabbri, Caiming Xiong, Shafiq Joty, and Chien-Sheng Wu. 2023.
\newblock Summedits: Measuring llm ability at factual reasoning through the lens of summarization.
\newblock In \emph{Proceedings of the 2023 conference on empirical methods in natural language processing}, pages 9662--9676.

\bibitem[{Li et~al.(2025)Li, Jiang, Huang, Beigi, Zhao, Tan, Bhattacharjee, Jiang, Chen, Wu et~al.}]{li2025generation}
Dawei Li, Bohan Jiang, Liangjie Huang, Alimohammad Beigi, Chengshuai Zhao, Zhen Tan, Amrita Bhattacharjee, Yuxuan Jiang, Canyu Chen, Tianhao Wu, and 1 others. 2025.
\newblock From generation to judgment: Opportunities and challenges of llm-as-a-judge.
\newblock In \emph{Proceedings of the 2025 Conference on Empirical Methods in Natural Language Processing}, pages 2757--2791.

\bibitem[{Li et~al.(2023)Li, Hovy, and Lau}]{li-etal-2023-summarizing}
Miao Li, Eduard Hovy, and Jey Lau. 2023.
\newblock \href {https://doi.org/10.18653/v1/2023.findings-emnlp.472} {Summarizing multiple documents with conversational structure for meta-review generation}.
\newblock In \emph{Findings of the Association for Computational Linguistics: EMNLP 2023}, pages 7089--7112, Singapore. Association for Computational Linguistics.

\bibitem[{Li et~al.(2024)Li, Lau, and Hovy}]{li-etal-2024-sentiment}
Miao Li, Jey~Han Lau, and Eduard Hovy. 2024.
\newblock \href {https://doi.org/10.18653/v1/2024.acl-long.547} {A sentiment consolidation framework for meta-review generation}.
\newblock In \emph{Proceedings of the 62nd Annual Meeting of the Association for Computational Linguistics (Volume 1: Long Papers)}, pages 10158--10177, Bangkok, Thailand. Association for Computational Linguistics.

\bibitem[{Liang et~al.(2024)Liang, He, Jiao, Wang, Wang, Wang, Yang, Shi, and Tu}]{liang2024encouraging}
Tian Liang, Zhiwei He, Wenxiang Jiao, Xing Wang, Yan Wang, Rui Wang, Yujiu Yang, Shuming Shi, and Zhaopeng Tu. 2024.
\newblock Encouraging divergent thinking in large language models through multi-agent debate.
\newblock In \emph{Proceedings of the 2024 conference on empirical methods in natural language processing}, pages 17889--17904.

\bibitem[{Liu et~al.(2024)Liu, Yu, Zhang, Zhang, and Xiao}]{liu2024automatic}
Xiaogeng Liu, Zhiyuan Yu, Yizhe Zhang, Ning Zhang, and Chaowei Xiao. 2024.
\newblock Automatic and universal prompt injection attacks against large language models.
\newblock \emph{arXiv preprint arXiv:2403.04957}.

\bibitem[{Liu et~al.(2023)Liu, Yao, Ton, Zhang, Guo, Cheng, Klochkov, Taufiq, and Li}]{liu2023trustworthy}
Yang Liu, Yuanshun Yao, Jean-Francois Ton, Xiaoying Zhang, Ruocheng Guo, Hao Cheng, Yegor Klochkov, Muhammad~Faaiz Taufiq, and Hang Li. 2023.
\newblock Trustworthy llms: a survey and guideline for evaluating large language models' alignment.
\newblock \emph{arXiv preprint arXiv:2308.05374}.

\bibitem[{Madaan et~al.(2023)Madaan, Tandon, Gupta, Hallinan, Gao, Wiegreffe, Alon, Dziri, Prabhumoye, Yang et~al.}]{madaan2023self}
Aman Madaan, Niket Tandon, Prakhar Gupta, Skyler Hallinan, Luyu Gao, Sarah Wiegreffe, Uri Alon, Nouha Dziri, Shrimai Prabhumoye, Yiming Yang, and 1 others. 2023.
\newblock Self-refine: Iterative refinement with self-feedback.
\newblock \emph{Advances in Neural Information Processing Systems}, 36:46534--46594.

\bibitem[{Naveed et~al.(2025)Naveed, Khan, Qiu, Saqib, Anwar, Usman, Akhtar, Barnes, and Mian}]{naveed2025comprehensive}
Humza Naveed, Asad~Ullah Khan, Shi Qiu, Muhammad Saqib, Saeed Anwar, Muhammad Usman, Naveed Akhtar, Nick Barnes, and Ajmal Mian. 2025.
\newblock A comprehensive overview of large language models.
\newblock \emph{ACM Transactions on Intelligent Systems and Technology}, 16(5):1--72.

\bibitem[{Orwig et~al.(1997)Orwig, Chen, Vogel, and Nunamaker}]{orwig1997multi}
R~Orwig, Hsinchun Chen, D~Vogel, and Jay~F Nunamaker. 1997.
\newblock A multi-agent view of strategic planning using group support systems and artificial intelligence.
\newblock \emph{Group Decision and Negotiation}, 6(1):37--59.

\bibitem[{Parcalabescu and Frank(2024)}]{parcalabescu2024measuring}
Letitia Parcalabescu and Anette Frank. 2024.
\newblock On measuring faithfulness or self-consistency of natural language explanations.
\newblock In \emph{Proceedings of the 62nd Annual Meeting of the Association for Computational Linguistics (Volume 1: Long Papers)}, pages 6048--6089.

\bibitem[{Park et~al.(2023)Park, O'Brien, Cai, Morris, Liang, and Bernstein}]{park2023generative}
Joon~Sung Park, Joseph O'Brien, Carrie~Jun Cai, Meredith~Ringel Morris, Percy Liang, and Michael~S Bernstein. 2023.
\newblock Generative agents: Interactive simulacra of human behavior.
\newblock In \emph{Proceedings of the 36th annual acm symposium on user interface software and technology}, pages 1--22.

\bibitem[{Piatti et~al.(2024)Piatti, Jin, Kleiman-Weiner, Sch{\"o}lkopf, Sachan, and Mihalcea}]{piatti2024cooperate}
Giorgio Piatti, Zhijing Jin, Max Kleiman-Weiner, Bernhard Sch{\"o}lkopf, Mrinmaya Sachan, and Rada Mihalcea. 2024.
\newblock Cooperate or collapse: Emergence of sustainable cooperation in a society of llm agents.
\newblock \emph{Advances in Neural Information Processing Systems}, 37:111715--111759.

\bibitem[{Rogelberg et~al.(2006)Rogelberg, Leach, Warr, and Burnfield}]{Rogelberg2006NotAM}
Steven Rogelberg, Desmond~J. Leach, Peter~B. Warr, and Jennifer~L. Burnfield. 2006.
\newblock \href {https://api.semanticscholar.org/CorpusID:16805873} {"not another meeting!" are meeting time demands related to employee well-being?}
\newblock \emph{The Journal of applied psychology}, 91 1:83--96.

\bibitem[{Romney et~al.(2025)Romney, Allen, and Heydarifard}]{ROMNEY202533}
Alexander~C. Romney, Joseph~A. Allen, and Zahra Heydarifard. 2025.
\newblock \href {https://doi.org/10.1016/j.bushor.2023.10.002} {Meeting load paradox: Balancing the benefits and burdens of work meetings}.
\newblock \emph{Business Horizons}, 68(1):33--43.

\bibitem[{Runco and Acar(2012)}]{runco2012divergent}
Mark~A Runco and Selcuk Acar. 2012.
\newblock Divergent thinking as an indicator of creative potential.
\newblock \emph{Creativity research journal}, 24(1):66--75.

\bibitem[{Saari(1995)}]{saari1995basic}
Donald~G. Saari. 1995.
\newblock \href {https://doi.org/10.1007/978-3-642-57748-2} {\emph{Basic Geometry of Voting}}, 1 edition.
\newblock Springer-Verlag Berlin Heidelberg.

\bibitem[{Sharma et~al.()Sharma, Tong, Korbak, Duvenaud, Askell, Bowman, DURMUS, Hatfield-Dodds, Johnston, Kravec et~al.}]{sharmatowards}
Mrinank Sharma, Meg Tong, Tomasz Korbak, David Duvenaud, Amanda Askell, Samuel~R Bowman, Esin DURMUS, Zac Hatfield-Dodds, Scott~R Johnston, Shauna~M Kravec, and 1 others.
\newblock Towards understanding sycophancy in language models.
\newblock In \emph{The Twelfth International Conference on Learning Representations}.

\bibitem[{Shi et~al.(2024)Shi, Ma, Liang, Diao, Ma, and Vosoughi}]{shi2024judging}
Lin Shi, Chiyu Ma, Wenhua Liang, Xingjian Diao, Weicheng Ma, and Soroush Vosoughi. 2024.
\newblock Judging the judges: A systematic study of position bias in llm-as-a-judge.
\newblock \emph{arXiv preprint arXiv:2406.07791}.

\bibitem[{Shin et~al.(2023)Shin, Koch, Lucero, Dalsgaard, and Mackay}]{10.1145/3544549.3573802}
Joon~Gi Shin, Janin Koch, Andr\'{e}s Lucero, Peter Dalsgaard, and Wendy~E. Mackay. 2023.
\newblock \href {https://doi.org/10.1145/3544549.3573802} {Integrating ai in human-human collaborative ideation}.
\newblock In \emph{Extended Abstracts of the 2023 CHI Conference on Human Factors in Computing Systems}, CHI EA '23, New York, NY, USA. Association for Computing Machinery.

\bibitem[{Shriberg et~al.(2004)Shriberg, Dhillon, Bhagat, Ang, and Carvey}]{shriberg-etal-2004-icsi}
Elizabeth Shriberg, Raj Dhillon, Sonali Bhagat, Jeremy Ang, and Hannah Carvey. 2004.
\newblock \href {https://aclanthology.org/W04-2319/} {The {ICSI} meeting recorder dialog act ({MRDA}) corpus}.
\newblock In \emph{Proceedings of the 5th {SIG}dial Workshop on Discourse and Dialogue at {HLT}-{NAACL} 2004}, pages 97--100, Cambridge, Massachusetts, USA. Association for Computational Linguistics.

\bibitem[{Song et~al.(2024)Song, Liu, Zhang, Zhang, Luo, Wang, Wu, and Wang}]{song2024adaptive}
Linxin Song, Jiale Liu, Jieyu Zhang, Shaokun Zhang, Ao~Luo, Shijian Wang, Qingyun Wu, and Chi Wang. 2024.
\newblock Adaptive in-conversation team building for language model agents.
\newblock \emph{arXiv preprint arXiv:2405.19425}.

\bibitem[{Sun et~al.()Sun, Yin, Li, Wu, Qiu, and Kong}]{suncorex}
Qiushi Sun, Zhangyue Yin, Xiang Li, Zhiyong Wu, Xipeng Qiu, and Lingpeng Kong.
\newblock Corex: Pushing the boundaries of complex reasoning through multi-model collaboration.
\newblock In \emph{First Conference on Language Modeling}.

\bibitem[{Tessler et~al.(2024)Tessler, Bakker, Jarrett, Sheahan, Chadwick, Koster, Evans, Campbell-Gillingham, Collins, Parkes, Botvinick, and Summerfield}]{doi:10.1126/science.adq2852}
Michael~Henry Tessler, Michiel~A. Bakker, Daniel Jarrett, Hannah Sheahan, Martin~J. Chadwick, Raphael Koster, Georgina Evans, Lucy Campbell-Gillingham, Tantum Collins, David~C. Parkes, Matthew Botvinick, and Christopher Summerfield. 2024.
\newblock \href {https://doi.org/10.1126/science.adq2852} {Ai can help humans find common ground in democratic deliberation}.
\newblock \emph{Science}, 386(6719):eadq2852.

\bibitem[{Tran et~al.(2025)Tran, Dao, Nguyen, Pham, O'Sullivan, and Nguyen}]{tran2025multi}
Khanh-Tung Tran, Dung Dao, Minh-Duong Nguyen, Quoc-Viet Pham, Barry O'Sullivan, and Hoang~D Nguyen. 2025.
\newblock Multi-agent collaboration mechanisms: A survey of llms.
\newblock \emph{arXiv preprint arXiv:2501.06322}.

\bibitem[{Wang et~al.(2023)Wang, Liu, Yue, Tang, Zhang, Jiayang, Yao, Gao, Hu, Qi et~al.}]{wang2023survey}
Cunxiang Wang, Xiaoze Liu, Yuanhao Yue, Xiangru Tang, Tianhang Zhang, Cheng Jiayang, Yunzhi Yao, Wenyang Gao, Xuming Hu, Zehan Qi, and 1 others. 2023.
\newblock Survey on factuality in large language models: Knowledge, retrieval and domain-specificity.
\newblock \emph{arXiv preprint arXiv:2310.07521}.

\bibitem[{Wang et~al.(2024)Wang, Li, Jiang, Tu, and Shi}]{Wang2024CraftingPAA}
Zheng Wang, Zhongyang Li, Zeren Jiang, Dandan Tu, and Wei Shi. 2024.
\newblock \href {https://api.semanticscholar.org/CorpusId:272986847} {Crafting personalized agents through retrieval-augmented generation on editable memory graphs}.
\newblock In \emph{Conference on Empirical Methods in Natural Language Processing}.

\bibitem[{Wei et~al.(2022)Wei, Wang, Schuurmans, Bosma, Xia, Chi, Le, Zhou et~al.}]{wei2022chain}
Jason Wei, Xuezhi Wang, Dale Schuurmans, Maarten Bosma, Fei Xia, Ed~Chi, Quoc~V Le, Denny Zhou, and 1 others. 2022.
\newblock Chain-of-thought prompting elicits reasoning in large language models.
\newblock \emph{Advances in neural information processing systems}, 35:24824--24837.

\bibitem[{Wu et~al.(2024)Wu, Bansal, Zhang, Wu, Li, Zhu, Jiang, Zhang, Zhang, Liu, Awadallah, White, Burger, and Wang}]{wu2024autogen}
Qingyun Wu, Gagan Bansal, Jieyu Zhang, Yiran Wu, Beibin Li, Erkang Zhu, Li~Jiang, Xiaoyun Zhang, Shaokun Zhang, Jiale Liu, Ahmed~Hassan Awadallah, Ryen~W White, Doug Burger, and Chi Wang. 2024.
\newblock \href {https://openreview.net/forum?id=BAakY1hNKS} {Autogen: Enabling next-gen {LLM} applications via multi-agent conversations}.
\newblock In \emph{First Conference on Language Modeling}.

\bibitem[{Yamaguchi(2021)}]{yamaguchi2021canvasvae}
Kota Yamaguchi. 2021.
\newblock Canvasvae: Learning to generate vector graphic documents.
\newblock \emph{ICCV}.

\bibitem[{Yu et~al.(2025)Yu, Meng, Zhou, Wang, Mao, Pan, Chen, Wang, Li, Zhang et~al.}]{yu2025survey}
Miao Yu, Fanci Meng, Xinyun Zhou, Shilong Wang, Junyuan Mao, Linsey Pan, Tianlong Chen, Kun Wang, Xinfeng Li, Yongfeng Zhang, and 1 others. 2025.
\newblock A survey on trustworthy llm agents: Threats and countermeasures.
\newblock In \emph{Proceedings of the 31st ACM SIGKDD Conference on Knowledge Discovery and Data Mining V. 2}, pages 6216--6226.

\bibitem[{Zhang et~al.(2025)Zhang, Yu, and Zhang}]{zhang2025systematic}
Haopeng Zhang, Philip~S Yu, and Jiawei Zhang. 2025.
\newblock A systematic survey of text summarization: From statistical methods to large language models.
\newblock \emph{ACM Computing Surveys}, 57(11):1--41.

\bibitem[{Zhang et~al.(2024{\natexlab{a}})Zhang, Zhang, Liu, Song, Wang, Krishna, and Wu}]{zhang2024offline}
Shaokun Zhang, Jieyu Zhang, Jiale Liu, Linxin Song, Chi Wang, Ranjay Krishna, and Qingyun Wu. 2024{\natexlab{a}}.
\newblock Offline training of language model agents with functions as learnable weights.
\newblock In \emph{Forty-first International Conference on Machine Learning}.

\bibitem[{Zhang and Xiong(2025)}]{zhang2025debate4math}
Shaowei Zhang and Deyi Xiong. 2025.
\newblock Debate4math: Multi-agent debate for fine-grained reasoning in math.
\newblock In \emph{Findings of the Association for Computational Linguistics: ACL 2025}, pages 16810--16824.

\bibitem[{Zhang et~al.(2024{\natexlab{b}})Zhang, Jin, Meng, Wang, and Tan}]{zhang2024comprehensive}
Yang Zhang, Hanlei Jin, Dan Meng, Jun Wang, and Jinghua Tan. 2024{\natexlab{b}}.
\newblock A comprehensive survey on process-oriented automatic text summarization with exploration of llm-based methods.
\newblock \emph{arXiv preprint arXiv:2403.02901}.

\bibitem[{Zhang et~al.(2024{\natexlab{c}})Zhang, Zhang, Liu, Fabbri, Liu, Kamoi, Lu, Xiong, Zhao, Radev, McKeown, and Zhang}]{zhang-etal-2024-fair}
Yusen Zhang, Nan Zhang, Yixin Liu, Alexander Fabbri, Junru Liu, Ryo Kamoi, Xiaoxin Lu, Caiming Xiong, Jieyu Zhao, Dragomir Radev, Kathleen McKeown, and Rui Zhang. 2024{\natexlab{c}}.
\newblock \href {https://doi.org/10.18653/v1/2024.naacl-long.187} {Fair abstractive summarization of diverse perspectives}.
\newblock In \emph{Proceedings of the 2024 Conference of the North American Chapter of the Association for Computational Linguistics: Human Language Technologies (Volume 1: Long Papers)}, pages 3404--3426, Mexico City, Mexico. Association for Computational Linguistics.

\bibitem[{Zhu et~al.(2025)Zhu, Yang, Bakker, Pentland, and Pei}]{zhu2025can}
Shenzhe Zhu, Shu Yang, Michiel~A Bakker, Alex Pentland, and Jiaxin Pei. 2025.
\newblock Can ai truly represent your voice in deliberations? a comprehensive study of large-scale opinion aggregation with llms.
\newblock \emph{arXiv preprint arXiv:2510.05154}.

\end{thebibliography}

\appendix
\clearpage
\section{Discussion}
In this section, we discuss the broader implications of our findings. We begin by examining the contributions of our user study procedure as a scalable method for evaluating AI systems for group ideation. We then consider the potential of \ours's architecture as a generalizable model for AI-facilitated team convergence beyond graphic design. Finally, we address the limitations of our work and outline promising directions for future research.

\subsection{User Study Procedure}
One of the primary contributions of this work is the evaluation procedure itself. Research into AI systems for team settings, particularly in creative domains, is often hampered by methodological challenges and logistical overhead in human-in-the-loop evaluation. A key advantage of our three-phase protocol (Annotate $\rightarrow$ Simulate $\rightarrow$ Re-evaluate) is its scalability, which addresses this important bottleneck in team settings beyond text-only domains.

While existing work relies on live, synchronous sessions with participants, making them time-consuming, expensive, and difficult to scale beyond a small number of test cases, our approach is asynchronous and simulation-driven. By collecting designers' detailed rankings and justifications upfront (Phase 1), we effectively treat their expert judgment as a reusable resource. Instead of requiring designers to be online for every system execution, our approach can compose nominal teams from the offline annotations, simulate team dynamics, and return to team members to evaluate groundedness and effectiveness from system-generated deliverables. This allowed us to run 100 test cases covering a larger experimental space (i.e., two team sizes, three iterations of debating-and-remixing) much more efficiently. We hope the community interested in group ideation \citep{10.1145/3544549.3573802} can benefit from the key ideas behind our reproducible protocol.


\subsection{AI as a Facilitator for Team Convergence in Other Creative Domains}

While \ours was implemented and evaluated within the domain of professional graphic design, its underlying architecture can be re-instantiated or extended to support team convergence in other creative domains. Advances in generative AI for video- or audio-editing, for example, pose interesting questions as to whether the positive findings we see in our studies would transfer to these other professional settings that similarly rely on group ideation. 



\subsection{Potential Risks}
One significant risk involves the privacy and security implications of creating high-fidelity proxy agents conditioned on sensitive personal data. Since \ours operates by encoding a participant's specific expressed preferences directly into a structured system prompt, there is an inherent risk that these digital proxies could inadvertently disclose more information than the user intended. For instance, while a user might strategically withhold certain views or ``hidden assumptions'' in a human-to-human setting, a proxy agent designed to ``advocate for the assigned preference'' might be manipulated via adversarial prompting~\citep{greshake2023not,liu2024automatic} or dialogue leaks~\citep{agarwal-etal-2024-prompt} to reveal private rationales, biases, or competitive strategies to other agents in the shared ``group chat'' environment.

\section{Future Work}

We are actively interested in exploring more hierarchical teams spanning different roles. Extending \ours to other creative domains is another exciting direction that we identify, as well as further exploring ``knobs'' in the underlying parameter space such as the number of system iterations. Another important avenue of future work relates to more dynamic persona modeling: for example, agents could be designed to dynamically update their preferences and rationales based on the ongoing dialogue, better mimicking human adaptability and belief revision~\citep{kobsa1989taxonomy,johansson2002user,del1998belief}.

A closely related challenge is resolving genuinely contradictory preferences within human teams. As will be discussed later in failure mode (Appendix~\ref{failure_mode}), when participants hold mutually exclusive positions, the remix phase can not resolve the conflict. This is not unique to our system; zero-sum disagreements are difficult even for human teams to settle without an explicit tie-breaking mechanism. We envision two complementary directions to address this. First, introducing hierarchical agent roles, where agents are assigned different levels of authority or domain expertise, could provide a principled resolution path: for instance, when an Art Director and a Junior Designer disagree on a core brand element, the system could defer to the more senior role, mirroring how professional teams navigate such impasses in practice. Recent work on hierarchical multi-agent systems~\citep{tran2025multi, han2024llm,guo2024embodied} suggests that layered control structures can improve coordination without sacrificing the benefits of distributed deliberation. Second, rather than forcing the system to resolve every conflict autonomously, unresolved trade-offs could be surfaced back to the human team as interactive decision points (e.g., parameter sliders, toggle options, or side-by-side variant previews), letting the group make the final call on genuinely irreconcilable differences. This hybrid approach would preserve \ours's ability to structure and externalize disagreement while keeping humans in control of decisions that require value judgments beyond what proxy agents can decide.

\section{Additional Results}

\subsection{Cost Analysis}

We analyze the cost of running \ours in this subsection.

For Task 1, we report the total cost for each task with GPT-4.1 as backbone. As shown in Table~\ref{tab:cost}, \ours is cheaper than both compute-matched baselines. MAD's generic agents receive the full summaries from all other agents, inflating token counts. Self-Refine regenerates the entire summary each iteration, compounding input length. In contrast, \ours agents contribute to a shared group chat where message history builds incrementally, keeping per-turn context lean and cost-effective.

\begin{table}[h]
\centering
\begin{tabular}{cc}
\hline
\textbf{Method} & \textbf{Avg. Cost ($10^{-2}$ USD)} \\
\hline
Direct      & 0.12 \\
CoT         & 0.19 \\
Self-Refine & 0.77 \\
MAD         & 1.18 \\
\hline
\textbf{TeamFusion} & \textbf{0.64} \\
\hline
\end{tabular}
\caption{Cost comparison between \ours and baselines on Task 1.}
\label{tab:cost}
\end{table}

For Task 2, candidate generation costs about \$0.90/scenario, and \ours revision costs about \$0.67/image, which is cost-effective given 42-50\% top-1 acceptance. In terms of latency, our live study shows that \ours reduces decision time from 18.0 to 12.4 minutes compared to free-form discussion, while also yielding higher representativeness and satisfaction scores.

\subsection{Ablations on Per Agent Turns}
\begin{figure}[t]
  \centering
  \includegraphics[width=\columnwidth]{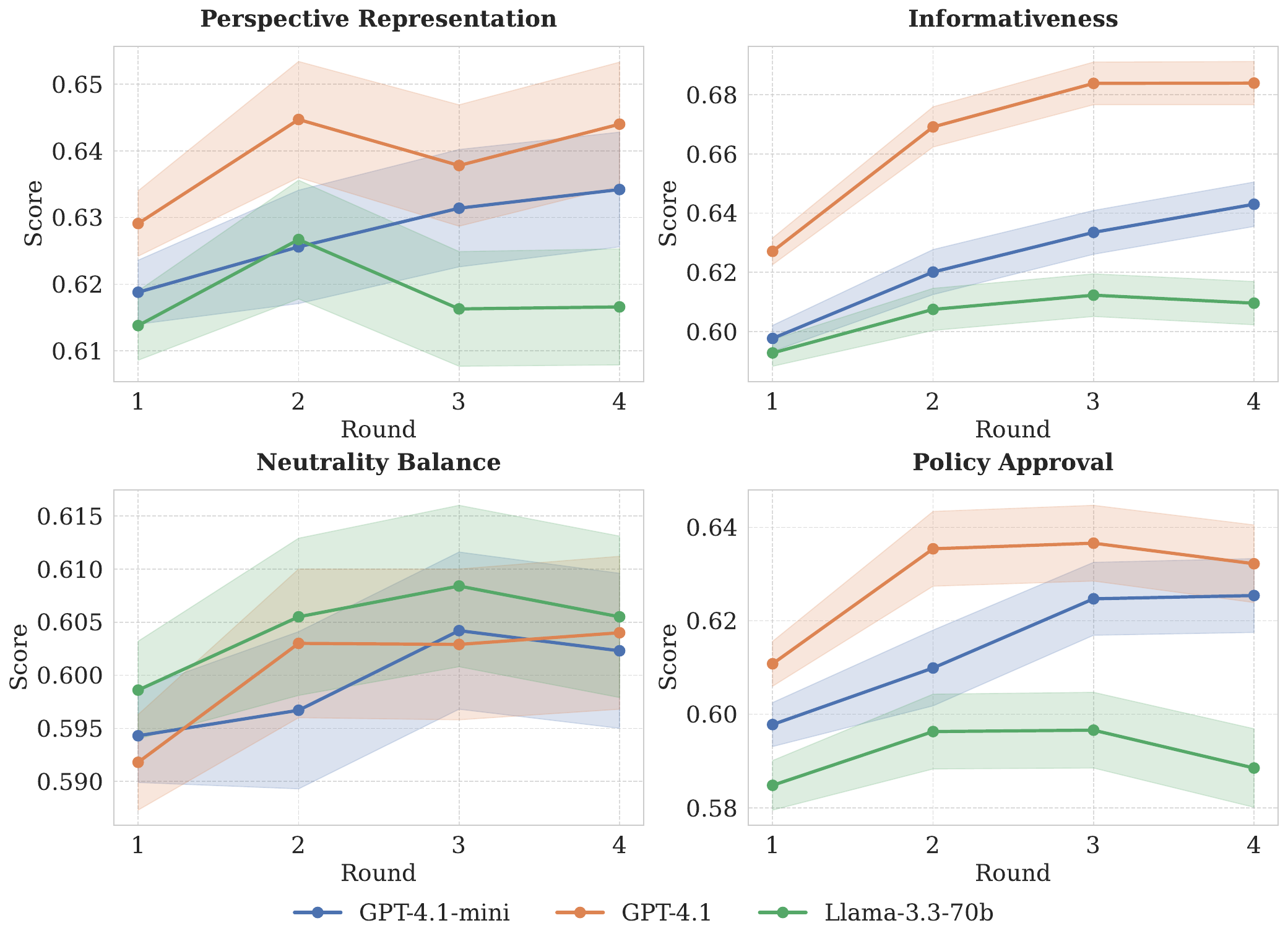}
  \caption{Ablations of per agent speaking turns on the four metrics.}
  \label{fig:discussion}
\end{figure}

We ablate on the number of discussion rounds in Task 1. We set the number of discussion rounds and observe \ours performance across the four metrics. As shown in Figure~\ref{fig:discussion}, scaling up the rounds of discussion can improve informativeness and neutrality. Stronger proprietary models like GPT-4.1-mini and GPT-4.1 benefits more from increasing discussion rounds, representativeness increases. This showcases that adding more discussion can make different voices heard for stronger models. For weaker model Llama, adding round from 3 to 4 yields a decrease in four metrics. We hypothesize that this is due to its worse long context processing capability.

\section{Case Study}
\label{case_study}
\subsection{Case Study on Successful Outcomes}
\begin{figure*}[t]
  \centering
  \includegraphics[width=0.9\textwidth]{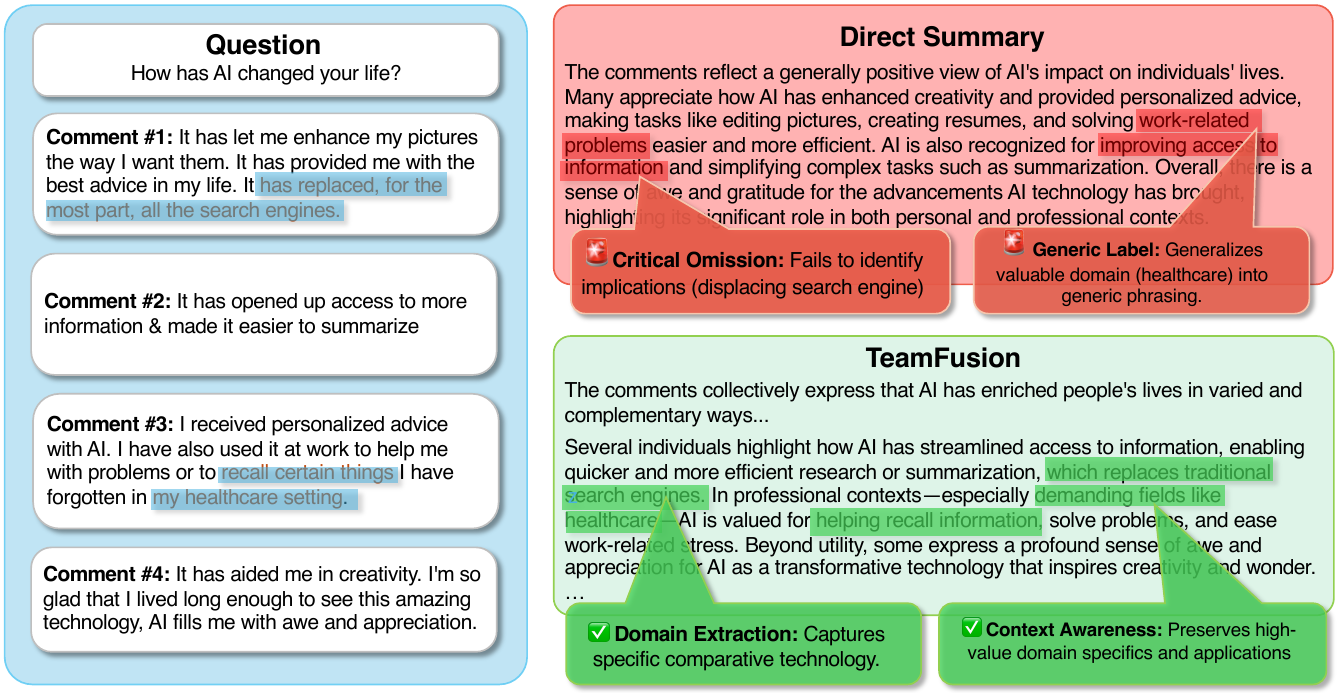}
  \caption{Case study in Task 1 comparing \ours and direct summary. We partially omit outputs from \ours to increase presentation focus.}
  \label{fig:case_study}
\end{figure*}

Figure~\ref{fig:case_study} illustrates how \ours better preserves fine-grained, participant-specific content than direct aggregation on a civic synthesis example. While the direct summary is fluent and captures the dominant positive sentiment (e.g., creativity, personalization, and efficiency), it noticeably homogenizes key details: it fails to carry forward Comment~\#1’s concrete comparative claim that AI has ``replaced'' traditional search engines, and it collapses Comment~\#3’s domain-specific experience (“my healthcare setting”) into generic ``work-related problems,'' obscuring where and why the tool is valuable. In contrast, \ours output retains these high-salience details in the final deliverable. This example highlights two practical advantages of \ours for generating deliverables: (i) minority or specialized experiences remain visible rather than averaged away, and (ii) concrete comparative statements and applications are maintained to support downstream interpretation and action. Overall, the case study qualitatively supports our quantitative gains on representativeness by showing that \ours more faithfully carries forward what each participant uniquely contributed, instead of compressing distinct voices into a generic narrative. 

\subsection{Case Study on Failure Mode}
\label{failure_mode}
While \ours demonstrates strong performance across our evaluations, we identify a recurring failure pattern that reveals a fundamental limitation of the framework. From our qualitative inspection, \ours underperforms when participants hold truly contradictory preferences about a specific topic. For example, in the visual design scenario, Designer A like a particular visual element while Designer B explicitly opposes it. Unlike cases where preferences differ in degree or emphasis, such zero-sum conflicts present an irreconcilable tension: any deliverable that satisfies one participant necessarily violates the other's stated constraint.

In these cases, the remix phase is forced to make a selection between the two conflicting viewpoints, leading to lower perceived representativeness for the team. This pattern is consistent with findings in the fair summarization literature, where aggregation over opposing stances risks producing outputs that no individual stakeholder endorses~\citep{zhang-etal-2024-fair}.

It is worth noting that this limitation is not unique to \ours. Zero-sum disagreements are inherently difficult even for human teams to resolve without an explicit tie-breaking mechanism such as authority, voting, or external arbitration~\citep{fisher1970decision, black1948rationale}. 

\section{Implementation Details of Task 1}

\subsection{Details of Represent}
We design a prompt that consists of goal of the discussion, conversation style constraints, and preference samples. Prompt content is available at Appendix~\ref{appdx:task2_represent}.


\subsection{Details of Remix}
The remixing agent is a text-based LLM that takes in the task context, original comments, the discussion transcript, and generates a structured summary. Prompt content can be found at Appendix~\ref{sec:remix_prompt}.

\subsection{Details of Iterative Revision}
After an initial summary has been generated, the summary becomes a shared group message content as part of the task context.
The agent group's collective goal changes to improve the summary to better reflect their individual standpoint. They engage in the structured discussion to achieve the goal. Finally, the remixing agent ingests the individual comments based on the previous summary and the debate transcript, and generates a refined summary of the comments. The newly generated summary then becomes the summary to improve upon in the next round, if any. 

\subsection{Hyperparameters}
We use three LLMs as backbones for both proxy agents and the remixing agent, i.e. Llama-3.3-70b, GPT-4.1-mini, GPT-4.1. We set the decoding temperature to 1. We set the number of per-debate turns to 1 in the main experiments. The agents are implemented and orchestrated by the AutoGen framework~\citep{wu2024autogen}.

\section{Implementation Details of Task 2}
\subsection{Details of Representation}
\label{init}

Based on recent works on LLM-based personalization~\citep{Chen2024LearningAMA,Huang2024SelectivePTA,Wang2024CraftingPAA,Kwon2023WhatWAA}, we leverage in-context learning to customize an LLM agent for the participant. We combine best practices from prior work to compose a layered prompt that makes an LLM adhere to a given designer's preferences and reliably act as individual Designer Agents in our multi-agent system:

\begin{itemize}
    \item \textbf{Overarching goal:} The first component of the prompt defines the overarching goal and the role of the agent. It establishes the agent's role as a 'design expert' and its goal is to reach consensus over potentially varying preferences.
    
    \item \textbf{Domain constraints:}
    Following the overarching goal, this component helps to concretize the ``best practices'' in communication that make the agent specifically natural-sounding and useful for design ideation. By prompting along more specific dimensions such as tone, language, and conciseness (e.g., \textit{``Mimic real designers' tone and language style... Avoid very long messages''}), we further align the agent's output space to a desired communication style. This is expected to make agents' outputs more natural-sounding when verified by human evaluators.

    \item \textbf{Role-playing definition:}
    This component is the core of the personalization, explicitly limiting the agent's behavior to role-playing the human designer. The instruction, \textit{``You are role-playing \{user\_name\}. Always respond from the following perspective and expertise,''} acts as a powerful anchor that instructs the model to forego its default, neutral stance and instead adopt the specific viewpoint, expertise, and potential biases of the individual it represents, becoming their debate proxy.
    
    \item \textbf{Few-shot preference examples:}
    Finally, to concretely ground the agent on a preference set, the prompt is completed with the designer's opinions over the option space. These opinions include both ordered preferences---each option's ranking from best to worst according to that designer---as well as brief \textit{ranking justifications} in natural language (e.g., \textit{Image 4: \{`rank': `1', `justification': ``It's bold and colorful, but feels more like a fashion brand than perfume. The bottle's squeezed in and doesn't really pop.''\}}). Considering the multi-modal nature of the input, which includes images and text, this component provides rich information for preference-grounding: the rank placements are explicit; the natural language justifications articulate explicit rationales; and even second-order preferences can be inferred from the image-text pair.
\end{itemize}

\subsection{Details of Discussion}
In Task 2, the overarching goal of the agent discussion is two-fold: (1) Reaching a consensus on the rankings of the six images according to adherence to the client brief and aesthetics; (2) Converging on the direction to improve based on the existing images.

\subsection{Details of Remixing Agent}
\label{sec:remix_detail}
Once all proxy designer agents have reached the number of turns per debate, \ours moves to translating the discussion outcomes into a revised image design. The remixing agent first consumes the entire chat history and extract two structured outputs: (1) The top-ranked options capturing the group's consensus, and (2) A set of remixing instructions, specifying which strengths from the top-ranked options to keep while addressing their weaknesses. The remixing agent then feeds the top-ranked options and the set of remixing instructions into a downstream image editing model. Importantly, this is a fundamental difference between a group brainstorming technique such as nominal group technique~\citep{10.1016/S0167-9236(00)00073-7}, which would apply voting at the end to obtain the top-ranked options, and \ours's integration of generative AI in a consensus-oriented manner, incorporating AI as a creative partner.

\subsection{Details of Iterative Revision}

If the team would like to run another iteration on \ours, the system is designed to narrow the scope on the top-ranked options and iteratively refine them. An initial option space with six designs is narrowed down to the three top-ranked options after the first debating iteration. A new remixed option is then added to this top three during remixing, finishing the first iteration with a top four. A second iteration would start from this top four, with the Designer Agents debating them for the same number of per-debate turns, narrowing down to a top two. A new remixed option would yield a top three at the end of the second iteration. Once iterative refinement is over, a final discussion yields the single best option for that run of \ours.

\subsection{Hyperparameters}
We use GPT-4o as the backbone for the agents, with decoding temperature set to 1. The agents are implemented and orchestrated by the AutoGen framework~\citep{wu2024autogen}. We leverage GPT-Image-1 for image remixing part of the Remixing Agent. Empirically, we have found that setting the number of per-debate turns to 2 allows agents in \ours to negotiate in-depth without repeating themselves sycophantically~\citep{sharmatowards}, with little practical utility, or going off-topic---so we have fixed this parameter to 2 in this task. We run \ours with 2 follow-up iterative revision rounds.

\section{User Study Details}

\subsection{Participants}
We recruited 9 professional graphic designers (6 female, 3 male) on the Upwork platform. Each designer had substantial experience in social media ad design as recorded on the platform (mean jobs completed = 76.56, $\sigma=95.18$), having successfully completed at the very least 10 projects. The compensation for participation varied by designer (mean = \$317.22, $\sigma=\$108.69$). Participants were anonymized and all procedures were approved by institutional review board.

\subsection{Scenario Construction}
\label{sec:data}
We constructed the scenario in a three-phase process. First, we sampled 70 social media ads from the Crello dataset~\cite{yamaguchi2021canvasvae}, filtering by the ``Facebook Ad'' and ``Instagram Ad'' categories. Second, we employed GPT-4o to reverse-engineer a hypothetical client brief for each ad image. Third, these client briefs and Crello ad images were sent into an image generation pipeline to create five new design variants for each setting. Specifically, the pipeline begins with a LLM planner that reads the image and the variation requirement, generates a plan to change certain aspects of the image. The plan is fed to a prompt writer LLM that consolidates the plan into a detailed instruction. GPT-Image-1 reads the instruction and the original image, then generates the image variant.

To validate the professional quality of client briefs and design options, we recruited two senior designers on Upwork who were native English speakers and had particularly extensive track records in dealing with real clients (311 and 520 completed projects). The designers scored all client briefs and design options on a 5-point Likert scale (1 = Completely Unrealistic, 5 = Fully Realistic)\footnote{Full instructions can be found in the supplemental materials.}, allowing us to select 50 high-quality social media ad scenarios with positive scores from both judges, each including a realistic client brief and six design options.

\subsection{Procedure}
\label{procedure}
The relative scarcity of research addressing AI systems in team settings, particularly in the graphic design domain, motivated us to plan a novel procedure for composing teams of professional graphic designers. At a high level, this procedure consists of assigning different designers to the same setting (i.e., a client brief + six options), collecting their initial preferences over the option space, composing nominal teams to compute team-wide preferences, simulating these teams on \ours, including \ours-generated designs when collecting their new individual preferences, and returning to the nominal teams to compute new team-wide preferences, thus evaluating \ours's contributions to the teams' final preferences.
In detail, this procedure was implemented in three phases:

\textbf{Phase 1: Initial designer annotation.} Designers reviewed each brief and its six associated design options.\footnote{Full instructions can be found in the supplemental materials.} They were asked to rank the options based on how well they---subjectively---felt that each option addressed the client brief. For each ranking decision, they provided short (i.e., 2-3 sentences) written justifications expressing their judging rationales. We ensured that each social media ad scenario (client brief + six options) received at least four independent annotations, with each designer participating in exactly 25 scenarios.

\textbf{Phase 2: \ours run.} For each of the 50 scenarios, we experimented with two team settings: (a) \textit{Full-Team}, where all available designer data for a scenario were used to initialize Designer Agents; and (b) \textit{Small-Team}, where random subsets of two designers were used to initialize Designer Agents. \ours was fully executed for both the Full- and Small-Team settings, through three iterations of debating-and-remixing, collecting the \ours-generated design at each iteration. As a result, we executed \ours $50\times2=100$ times, and collected a total of $100\times 3=300$ \ours-generated options. From each execution, we also collect the simulated comments (per Section \ref{iter}) that each Designer Agent makes for the three \ours-generated options---all unseen by the original annotators.

\textbf{Phase 3: Designer re-evaluation.} For each of the 100 team settings, we collected the team's top three options using Borda count~\cite{saari1995basic} from their initial rankings. We then mixed the three \ours-generated options with the initial top three, for a new set of six options. By ranking \ours's outputs vs. the initial top-ranked options, we can measure if---and to what extent---\ours is able to modify the teams' top-ranked preferences. After being exposed to the unseen options in the re-ranking task, designers then score on a 5-point Likert scale (1 = Strongly Disagree, 5 = Strongly Agree) their agreement with the simulated comments, measuring if---and to what extent---they feel represented by their proxy agents.

\subsection{Participant Task Instructions}
\begin{figure}
    \centering
    \includegraphics[width=\linewidth]{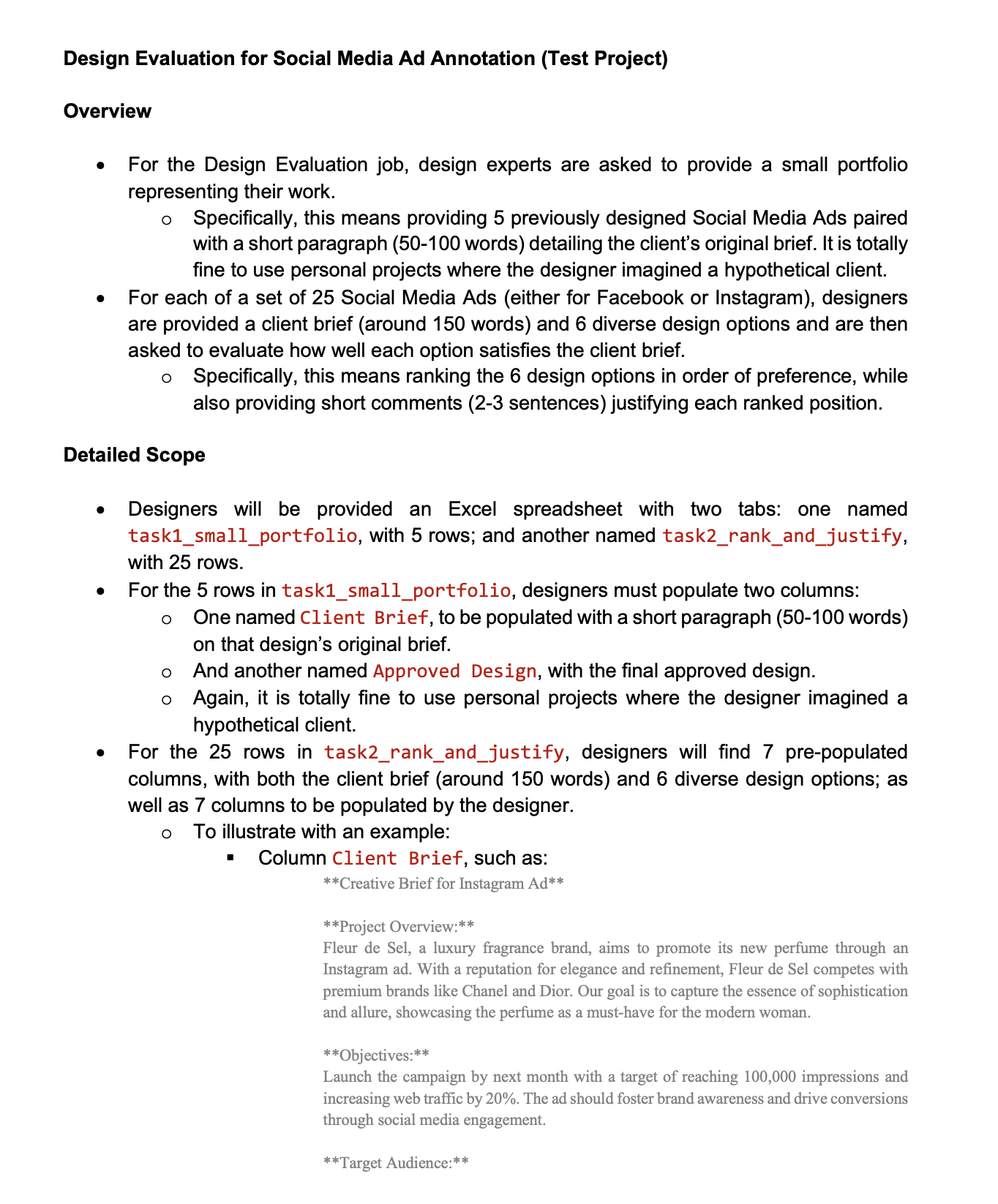}
    \caption{Task instruction part 1.}
    \label{fig:task_pt1}
\end{figure}

\begin{figure}
    \centering
    \includegraphics[width=\linewidth]{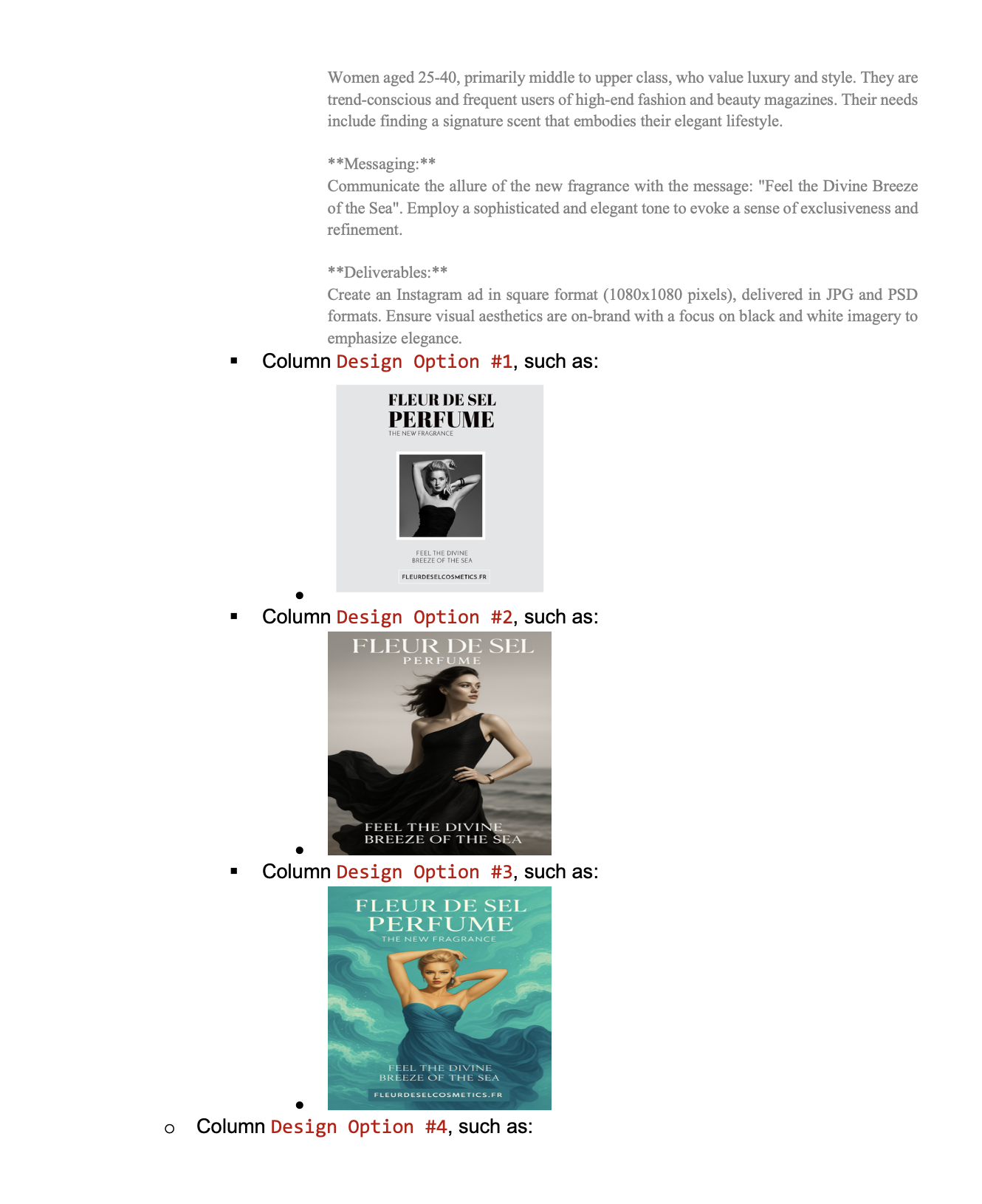}
    \caption{Task instruction part 2.}
    \label{fig:task_pt2}
\end{figure}

\begin{figure}
    \centering
    \includegraphics[width=\linewidth]{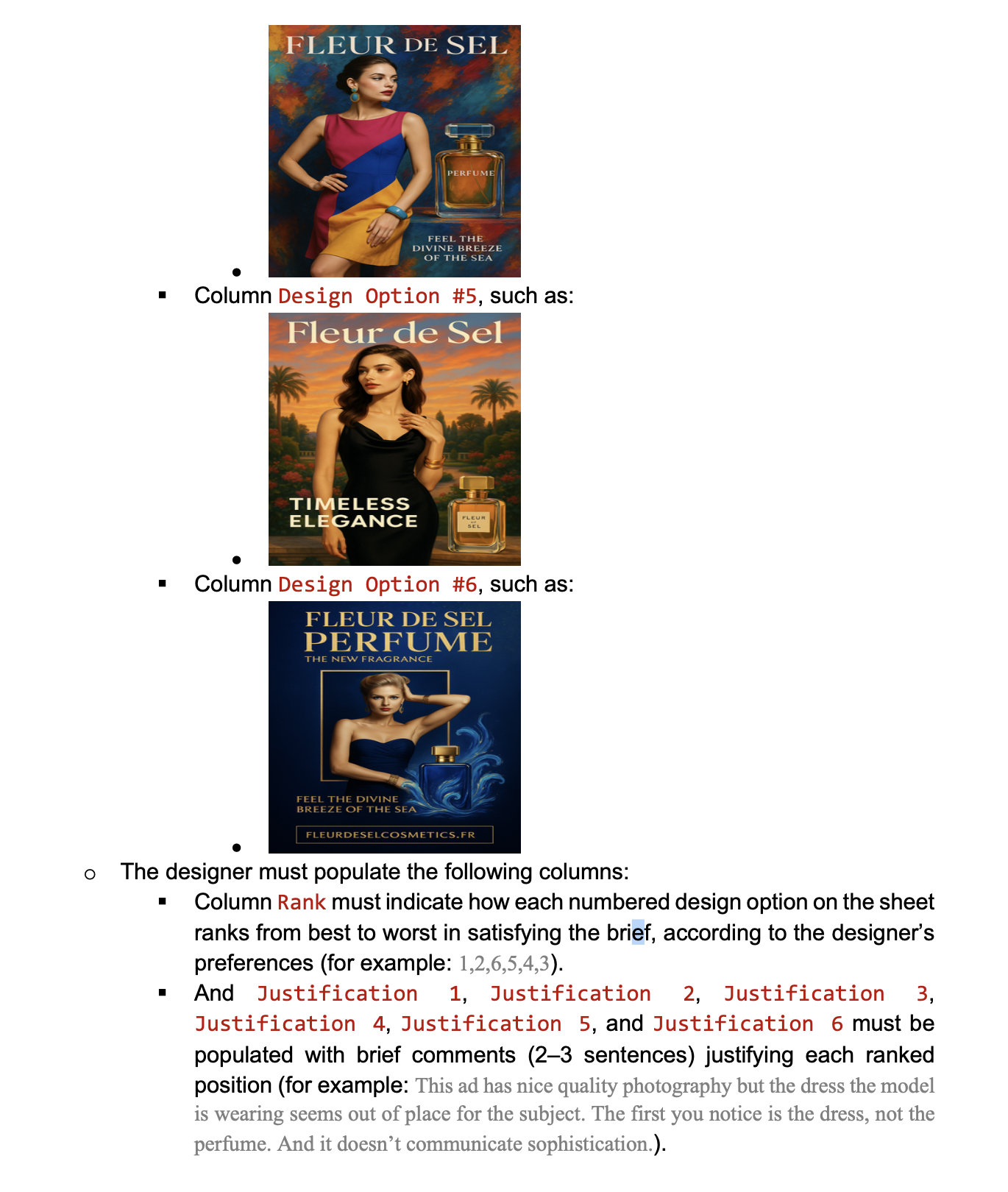}
    \caption{Task instruction part 3.}
    \label{fig:task_pt3}
\end{figure}

We present the task briefing to the participants in Figure~\ref{fig:task_pt1}, \ref{fig:task_pt2} and~\ref{fig:task_pt3}.


\section{Live User Study Details}
\label{app:live-study}

\subsection{Goal}
We conducted a small live study to evaluate \ours in an end-to-end collaborative design workflow where team members (i) create initial candidate ad thumbnails using generative tools, (ii) express individual preferences via rankings and rationales, and (iii) collaboratively converge on a final selection through either \ours or through free-form discussion and revision. The study focuses on decision-process outcomes and participant-reported experience.

\subsection{Experiment Protocol}
We recruited 6 participants and formed 2 teams of 3. Each team completed two ad-brief tasks (Brief A and Brief B). We \textbf{counterbalanced condition order} at the team level:
(i) Team 1 used \ours on Brief A and the baseline on Brief B; 
(ii) Team 2 used the baseline on Brief A and \ours on Brief B.
This controls for brief-specific difficulty and order effects.

\subsection{Materials}
\paragraph{Ad briefs.}
We prepared two ad briefs following the same protocol as~\ref{sec:data}.

\paragraph{Reference gallery.}
For each brief, we provided a gallery of 10 image thumbnails as references for generation.
This is to seed stylistic directions and reduce cold-start variance in what participants create.

\paragraph{Generative tools.}
All participants had access to the same GenAI editing and generation tools, specifically GPT image editing and Nano-Banana. 

\subsection{Conditions}
\paragraph{\ours}
After participants provided their initial rankings and short rationales, we constructed one proxy agent per participant using these preference signals. \ours then produced:
(1) a structured summary of agreements, disagreements, and key trade-offs;
(2) two revised candidate thumbnails.
Participants could optionally decide up to 3 additional refinement rounds by providing brief feedback (e.g., “strengthen product visibility”).

\paragraph{Free-form discussion}
Participants coordinated through Zoom to discuss freely. They posted revisions through a private channel on Discord. They could use the tools to propose revised thumbnails and post them to the channel. 

\subsection{Procedure}
Each brief followed the same phases:

\paragraph{Phase 0: Training (10 minutes)}
Participants first get aquainted with each other, then completed a short tutorial on the interface and tools by looking at admin-provided video demo.

\paragraph{Phase 1: Individual creation (15 minutes)}
Using the reference gallery and GenAI tools, each participant produced two initial candidate thumbnails. Participants uploaded their candidates to a shared board.

\paragraph{Phase 2: Individual preference elicitation (10 minutes).}
Participants independently ranked all initial candidates from best to worst for the brief and provided short justifications describing their key criteria. 

\paragraph{Phase 3: Collaborative revision and convergence (20 minutes).}
Participants then completed one of the two conditions:

\begin{itemize}
  \item \textbf{\ours:} We ran \ours using the Phase 2 evidence, producing revised designs and a structured trade-off summary. Participants reviewed the output and either (i) stopped and moved to Phase 4, or (ii) provided brief feedback to trigger at most 3 additional refinement rounds.
  \item \textbf{Free-form discussion and revision:} Participants read each other's preference, and discussed freely in the voice channel and posted revisions they generated. The team stopped when time elapsed or when they agreed on a final candidate set.
\end{itemize}

\paragraph{Phase 4: Post-task questionnaire (1 minute).}
After the study, participants answered three 1--5 Likert items (1=strongly disagree, 5=strongly agree):
\begin{itemize}
  \item \textbf{Q1 (Representativeness):} The final output reflected my key preferences and reasoning.
  \item \textbf{Q2 (Clarity):} It was clear what the main agreements, disagreements and trade-offs were and why.
  \item \textbf{Q3 (Outcome satisfaction):} I am satisfied with the final design decision our team reached.
\end{itemize}

After completing both briefs, participants answered a preference question: \emph{“Which workflow would you choose for similar tasks?”} (\ours / Free-form discussion).

\subsection{Study Interface}

\begin{figure}[htbp]
  \centering
  \includegraphics[width=\columnwidth]{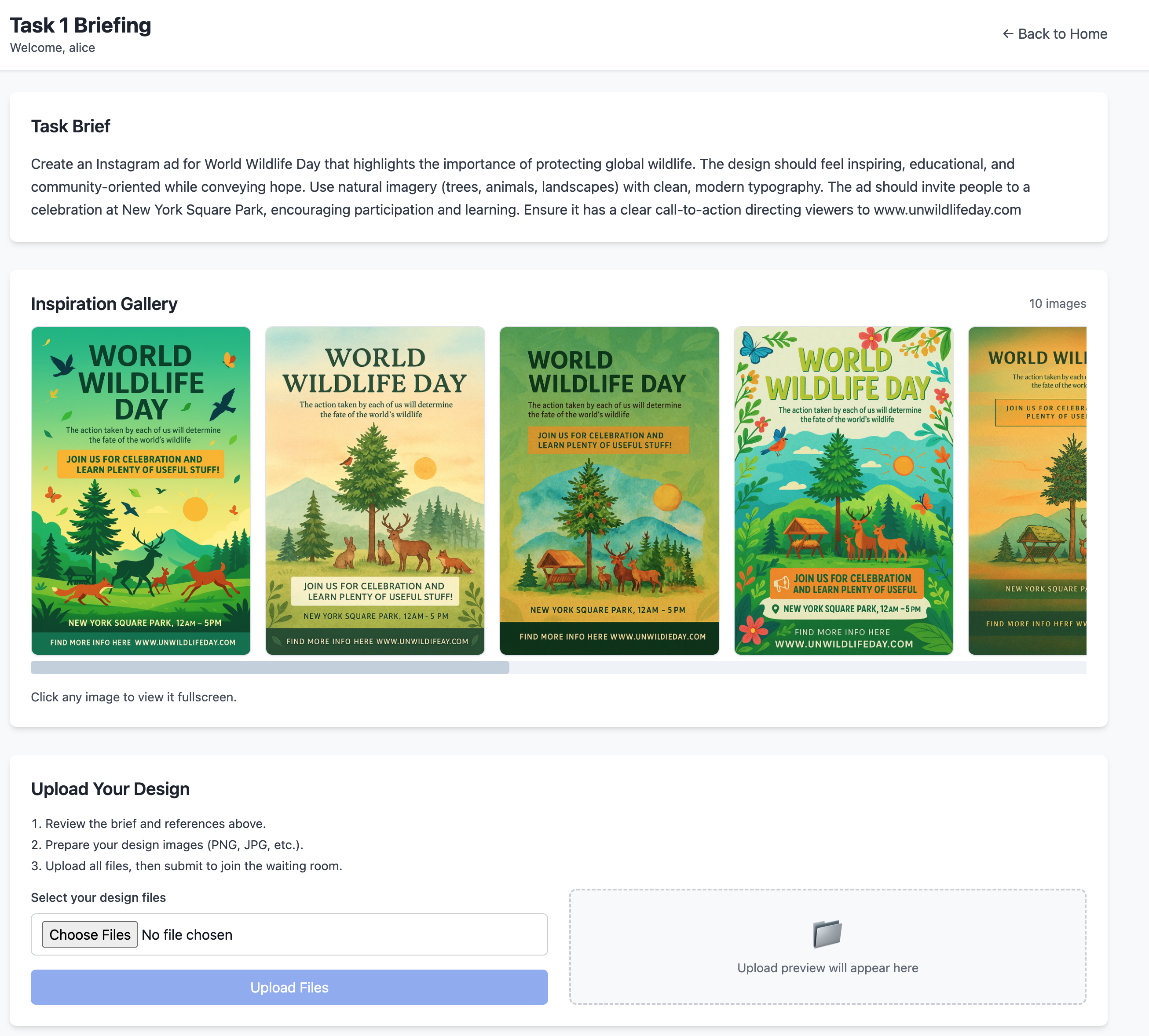}
  \caption{Screenshot of the live user study interface. This is the user upload design page. 
  }
  \label{fig:live1}
  
\end{figure}

\begin{figure}[htbp]
  \centering
  \includegraphics[width=\columnwidth]{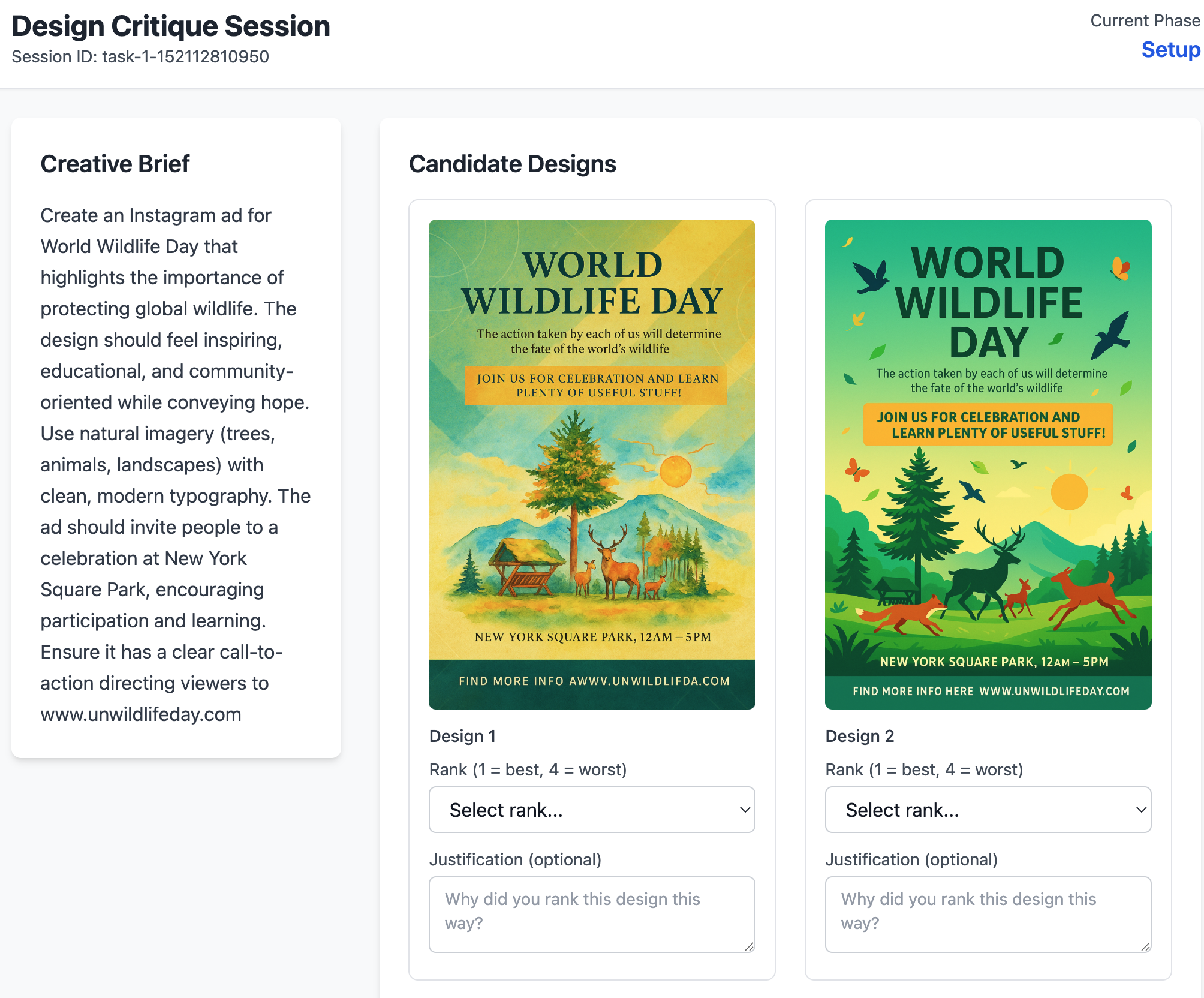}
  \caption{Screenshot of the live user study interface. This is the user critique design page. 
  }
  \label{fig:live2}
  
\end{figure}

We present the screen shots of the UI used in live user study in Figure~\ref{fig:live1}, \ref{fig:live2}. The UI enables users to upload their designs, writing critiques, reading each other's preferences, and monitoring \ours output. The screenshot uses mock data for visualization purposes.

\section{Detailed Explanation of Metrics}
In Table~\ref{tab:task2_metrics_decision_support}, we describe the four metrics and explain why they matter for evaluating open-ended team decisions.

\begin{table*}[t]
\centering
\small
\begin{tabular}{p{2.6cm} p{5.2cm} p{7cm}}
\toprule
\textbf{Metric} & \textbf{What it measures in this task} & \textbf{Why it matters for open-ended team decisions} \\
\midrule
Representativeness &
Whether the summary covers the range of participant viewpoints and attributes key reasons/claims to the underlying comments (i.e., avoids viewpoint erasure). &
Open-ended decisions require a deliverable that participants recognize as ``their'' perspectives being present.
High representativeness reduces minority suppression and makes disagreement auditable rather than implicitly averaged away. \\

Informativeness &
Whether the summary preserves concrete, decision-relevant content (rationales, constraints, trade-offs, edge cases), instead of generic paraphrases. &
Teams need a deliverable that supports action: identifying what information would change a decision, what trade-offs are being made, and what constraints are binding.
Higher informativeness makes the deliverable usable for follow-up discussion and planning. \\

Neutrality &
Whether the summary maintains a balanced, non-editorial tone and avoids injecting the summarizer's own stance. &
In civic/team settings, the deliverable often serves as shared ground for discussion.
Neutrality helps prevent the system from ``deciding'' for the group via framing effects, preserving legitimacy and trust in the deliverable. \\

Policy approval &
Whether the deliverable supports downstream acceptability/action (e.g., framing options in a way that is coherent, feasible, and aligned with the decision question). &
Open-ended deliverables are judged not only by coverage, but by whether they help a team move forward.
Policy approval captures whether the synthesized output is decision-oriented rather than merely descriptive. \\
\bottomrule
\end{tabular}
\caption{Evaluation dimensions for civic comment synthesis and their decision-support interpretation. We follow DeliberationBank’s four metrics and interpret them as complementary requirements for open-ended team deliverables.}
\label{tab:task2_metrics_decision_support}
\end{table*}

\section{AI Usage Disclosure}
AI assistants were used to help polish the manuscript’s wording and readability and to assist with drafting code snippets. All AI-assisted outputs were reviewed, verified, and edited by the authors.

\section{Prompts}
In this section, we list all the prompts used in the experiments.
\subsection{Task 1 Prompt}

\subsubsection{Prompt for Proxy Agent}
\label{appdx:task2_represent}
{
\ttfamily
\small
Your name is \{name\}. You are a participant in a public deliberation discussion. Your role is to advocate for and discuss the perspective expressed in your assigned comment.

\#\# Discussion Context
**Question:** \{question\}

**Your Assigned Comment:**
\{comment\}

\#\# Your Role and Instructions

1. **Understand and Hold Your Position**: Carefully read and internalize the viewpoint expressed in your comment. This represents your perspective in this discussion. Stay true to the sentiment and reasoning of your assigned comment. 

2. **Advocate Effectively**: 

   - Express the key points and reasoning behind your position
   
   - Always speak in concise and at most 2 paragraphs. Go straight to the core point.
   
   - Avoid adding any additional personal information or experience into discussion aside from given comments.

3. **Engage Constructively**:

   - Listen to and acknowledge other participants' viewpoints
   
   - Identify common ground where it exists
   
   - Respectfully challenge points you disagree with, using reasoning and evidence

4. **Contribute to Comprehensive Understanding**: Help ensure that your perspective is clearly understood and represented in the broader discussion, especially if it represents a minority or less common viewpoint.

Remember: The goal is not to "win" the debate, but to ensure all perspectives—including minority opinions—are thoroughly heard, understood, and considered in the final summary of the deliberation.

}

\subsubsection{Prompt for Remix Agent}
\label{sec:remix_prompt}
{
\ttfamily
\small
You are summarizing a collection of comments for a deliberation question: \{question\}. You will first receive the comments. Then, a discussion between people who wrote the comments will follow. You must focus on comprehensively summarizing the comments and use the discussion to better understand the viewpoints of the comments. Please do not mention the total number of comments. Do not refer to any specific comment in the summary. If you need to provide statistical information, use percentages instead of absolute numbers.
    
    Here are the comments: 
    
    \{comments\_str\}

    Here is the discussion, use it to better understand the comments:
    
    \{history\_str\}
}

\subsection{Task 2 Prompt}

\subsubsection{System Prompt for Proxy Agent}
The system prompt for Designer Agent has been presented and discussed in Section~\ref{init}.

\texttt{
\small
You are a design expert participating in a discussion about design images. You have been given specific
preferences over the designs and will discuss them with other experts to reach consensus. Advocate for
your preferred designs while being open to other perspectives. Mimic how real designers’ tone and
language style to write to each other. Be concise and to the point. You are roleplaying as \{user\_name\}.
Always respond from the following perspective and expertise. Attached are the images paired with the
justification, roleplay as if this is your preference.
\{formatted\_preference\}
}

\subsubsection{Prompt for Discussion}

{
\ttfamily
\small
Welcome to the design discussion! Each of you has seen the same set of images 
but may have different preferences. Please discuss efficiently and work toward consensus on:

1. RANKING: Establish a ranked list of images from best to worst, 
considering both aesthetic appeal and alignment with the creative brief.

2. DESIGN IMPROVEMENT: Discuss how to enhance and combine the best elements from top 3 performing images. Consider:

   - Primary composition and layout structure from the strongest images
   
   - Visual elements that should be integrated or refined
  
   - Color schemes and typography that work best
   
   - Specific adjustments needed to balance different concerns

3. SYNTHESIS: Develop a cohesive approach that merges strengths from the top 3 performing images 
while addressing any weaknesses identified in the discussion. When you propose changes to improve, ground the instructions on top 3 performing images.

Share your reasoning and be open to different perspectives as you work toward 
both a final ranking and concrete design improvement directions.

\vspace{2pt}
Here is the creative brief for the task:

\{brief\}
}

\subsubsection{Prompt for Remixing Agent}

{
\ttfamily
\small

You are a design summarization expert analyzing a roundtable discussion between design experts about image variants. Your task is to carefully read through the entire conversation and extract two key outputs:

1. **FINAL RANKING**: Identify the consensus ranking of images from best to worst

2. **EDITING DIRECTIONS**: Extract specific instructions for creating an improved design by combining elements from different images

\#\# Analysis Instructions:

- Start from the END of the conversation and work backwards - the most recent messages contain the final consensus and should be given the highest priority. Early messages may contain initial disagreements or positions that were later changed.

- Focus on extracting the ultimate agreements on rankings and specific design recommendations that emerged at the conclusion of the discussion.

\#\# Requirements for editing\_directions string:

Write detailed instructions as if directing an AI image editing model. It should include the following fields, if mentioned. If the discussion does not touch on the relevant field, don't include the field in your instruction.

1. **Primary Composition**. Example templates include:

Use the overall layout and structure from Image [number], specifically [describe the compositional elements, positioning, or arrangement].

2. **Visual Elements Integration**. Example templates include:

   - Incorporate [specific visual element] from Image [number], such as [detailed description]
   
   - Add [specific design feature] from Image [number], particularly [detailed description]
   
   - Include [specific element] from Image [number], focusing on [detailed description]

3. **Color and Typography Refinements**. Example templates include:

   - Adopt the [color scheme/typography style] from Image [number], specifically [details]
   
   - Modify [specific aspect] using the approach seen in Image [number]

4. **Final Adjustments**. Example templates include:

   - Ensure [specific requirement based on discussion]
   
   - Balance [specific concern raised in discussion]
   
   - Maintain [specific positive aspect mentioned]
   
\#\# Important Guidelines:

- Always reference images by their specific numbers (Image 1, Image 2, etc.)

- Be concrete and specific about visual elements (colors, positioning, typography, objects, etc.)

- Avoid vague language - use precise descriptions

- Focus on actionable instructions that an image editing AI could follow

- Only include elements and instructions that were actually discussed and agreed upon, never add your own novel thoughts

- If no clear consensus was reached, state this explicitly

\#\# Example 'editing\_directions':
"Incorporate the bold red CTA button from Image 3, positioning it in the lower-right corner as seen in Image 1, while maintaining the clean white background and centered product placement from Image 5."

Remember: Your output should be directly usable by downstream image editing systems, so precision and specificity are crucial.

\#\# Output Format:

You must output your analysis in the following JSON structure:

\begin{verbatim}
```json
{
  "final_ranking": [
    {
      "rank": 1,
      "image_number": <image_number>,
      "reason": "<reason_for_ranking>"
    },
    {
      "rank": 2,
      "image_number": <image_number>,
      "reason": "<reason_for_ranking>"
    },
    {
      "rank": 3,
      "image_number": <image_number>,
      "reason": "<reason_for_ranking>"
    }
  ],
  "editing_directions": "<instructions>"
}

```
\end{verbatim}

}

\end{document}